\documentclass[manuscript, nonacm]{acmart}

\AtBeginDocument{%
  \providecommand\BibTeX{{%
    \normalfont B\kern-0.5em{\scshape i\kern-0.25em b}\kern-0.8em\TeX}}}

\newcommand{\TOM}[0]{$\textit{TOM}$} 
\newcommand{\TheOtherMe}[0]{\textit{The Other Me}} 
\newcommand{\Jerry}[0]{$\textit{Jerry}$} 

\newif\ifcomment
\commentfalse

\newif\ifrevision
\revisiontrue

\ifcomment

\ifrevision

\newcommand{\hide}[1]{}
\newcommand{\note}[1]{}
\newcommand{\added}[1]{\textcolor[rgb]{1,0,0}{#1}}

\newcommand{\cut}[1]{}

\newcommand{\todo}[1]{}

\newcommand{\deleted}[1]{\textcolor[rgb]{0.6,0.6,0.6}{{#1}}}

\newcommand{\temporary}[1]{}

\newcommand{\nuwan}[1]{}
\newcommand{\david}[1]{}
\newcommand{\zsd}[1]{}
\newcommand{\sure}[1]{}

\else

\newcommand{\hide}[1]{}
\newcommand{\note}[1]{\textcolor{blue}{<< #1 >>}}
\newcommand{\added}[1]{\textcolor[rgb]{0.1, 0.56, 1}{#1}}

\newcommand{\cut}[1]{\textcolor[rgb]{0.5,0.5,0.5}{CUT: #1}}

\newcommand{\todo}[1]{\textcolor{red}{TODO: #1}}

\newcommand{\deleted}[1]{\textcolor[rgb]{0.7,0.7,0.7}{{#1}}}

\newcommand{\temporary}[1]{\textcolor[rgb]{0.9,0.9,0.9}{{#1}}}

\newcommand{\nuwan}[1]{\textcolor{blue}{NUWAN: #1}}
\newcommand{\david}[1]{\textcolor[rgb]{0.5,0.8,0}{DAVID: #1}}
\newcommand{\zsd}[1]{\textcolor[rgb]{0.5,0.1,0.8}{ZSD: #1}}
\newcommand{\sure}[1]{\textcolor[rgb]{0,0.8,0.4}{SURE: #1}}

\fi

\else
\newcommand{\hide}[1]{}
\newcommand{\note}[1]{}
\newcommand{\added}[1]{#1}

\newcommand{\cut}[1]{}

\newcommand{\todo}[1]{}

\newcommand{\deleted}[1]{}

\newcommand{\temporary}[1]{}

\newcommand{\nuwan}[1]{}
\newcommand{\david}[1]{}
\newcommand{\zsd}[1]{}
\newcommand{\sure}[1]{}

\fi

\renewcommand{\quote}[1]{``#1''}

\newcommand{\RIa}[0]{$\textbf{C1}_{a}$}
\newcommand{\RIb}[0]{$\textbf{C1}_{b}$}
\newcommand{\RIc}[0]{$\textbf{C1}_{c}$}

\newcommand{\RIIa}[0]{$\textbf{C2}_{a}$}
\newcommand{\RIIb}[0]{$\textbf{C2}_{b}$}
\newcommand{\RIIc}[0]{$\textbf{C2}_{c}$}

\newcommand{\RIIIa}[0]{$\textbf{C3}_{a}$}
\newcommand{\RIIIb}[0]{$\textbf{C3}_{b}$}
\newcommand{\RIIIc}[0]{$\textbf{C3}_{c}$}

\usepackage{subcaption}
\usepackage{multirow}
\usepackage{graphicx}
\usepackage{xcolor}
\usepackage{enumitem}
\usepackage{xspace}
\usepackage{booktabs}

\usepackage[nomessages]{fp}
\usepackage{soul} 

\newlength\maxlen

\def\databarlength{xx.xx} 
\begin{document}

\title[\TheOtherMe{}]{\TOM{}: A Development Platform For Wearable Intelligent Assistants}


\author{Nuwan Janaka}
\email{nuwanj@u.nus.edu}
\orcid{0000-0003-2983-6808}

\affiliation{%
  \institution{Synteraction Lab,}
  \institution{Smart Systems Institute, National University of Singapore}  
  \country{Singapore}
}

\author{Shengdong Zhao}
\authornote{Corresponding Authors.}
\email{shengdong.zhao@cityu.edu.hk}
\orcid{0000-0001-7971-3107}

\affiliation{%
\institution{Synteraction Lab,}
\institution{School of Creative Media \& Department of Computer Science, City University of Hong Kong}
\city{Hong Kong}
  \country{China}
}

\author{David Hsu}
\authornotemark[1]
\email{dyhsu@comp.nus.edu.sg}
\orcid{0000-0002-2309-4535}

\affiliation{%
  \institution{School of Computing,} 
  \institution{Smart Systems Institute, National University of Singapore}
  \country{Singapore}
}

\author{Sherisse Tan Jing Wen}
\email{sherisse_tjw@u.nus.edu}
\orcid{0009-0003-7078-2642}

\affiliation{%
  \institution{School of Computing, National University of Singapore} 
  \country{Singapore}
}

\author{Koh Chun Keat}
\email{idmkck@nus.edu.sg}
\orcid{0000-0001-9879-5419}

\affiliation{%
  \institution{Smart Systems Institute, National University of Singapore}
  \country{Singapore}
}

\renewcommand{\shortauthors}{Janaka et al.}


\begin{abstract}

Advanced digital assistants can significantly enhance task performance, reduce user burden, and provide personalized guidance to improve users' abilities. However, the development of such intelligent digital assistants presents a formidable challenge. To address this, we introduce \TOM{}, a conceptual architecture and software platform (\added{\url{https://github.com/TOM-Platform}}) designed to support the development of intelligent wearable assistants that are contextually aware of both the user and the environment. This system was developed collaboratively with AR/MR researchers, HCI researchers, AI/Robotic researchers, and software developers, and it continues to evolve to meet the diverse requirements of these stakeholders. \TOM{} facilitates the creation of intelligent assistive AR applications for daily activities and supports the recording and analysis of user interactions, integration of new devices, and the provision of assistance for various activities. Additionally, we showcase several proof-of-concept assistive services and discuss the challenges involved in developing such services.

\end{abstract}



\maketitle


\section{Introduction}

With advancements in Machine Learning (ML) and Artificial Intelligence (AI) technologies, digital assistants have become integral to everyday life. These include voice assistants like Siri, Alexa, or Google, and some even support visual modality (e.g., \cite{noauthor_ai_2024, rabbit2023}), enabling rich inputs and outputs. Similar to science fiction, advanced digital assistants can practically aid users in performing both familiar and new tasks, reduce task load and errors, and enhance task performance \cite{castelo_argus_2023}. Moreover, these assistants offer personalization, optimizing support for individual needs and broadening accessibility.

Despite existing interaction paradigms such as Heads-Up Computing \cite{zhao_heads_up_2023} and Dynamicland \cite{victor_dynamicland_2018} aiming to realize digital assistance in daily activities, several challenges persist. These include a lack of understanding of the required system capabilities, development guidance, and a platform for rapid development of such assistance (see Sec~\ref{sec:related_work}-\ref{sec:tom} for details).

To tackle these challenges, we introduce \TOM{}, an intelligent wearable assistive system developed by identifying user, researcher, and developer needs. \TOM{} facilitates the creation and analysis of assistive applications, enabling context and user understanding and supporting multimodal interactions with AR/MR devices and ML/AI technologies. Through developing several proof-of-concept services (e.g., running coach assistance, translation, and querying assistance), we have identified challenges in different daily activities and the necessary future improvements for the \TOM{} system to support envisioned usage scenarios.

The contributions of this paper are twofold: 1) Identifying the required capabilities for an intelligent wearable assistive system and developing a conceptual architecture; 2) Creating the \TOM{} system platform to enable the development of assistive services for various activities and demonstrating its application in several use cases.



\section{Related Work}
\label{sec:related_work}

Our work, which envisions an intelligent wearable assistive system, is related to context-aware and assistive Augmented Reality (AR) systems.

\subsection{Context-Aware Systems}
\label{sec:related_work:context_aware}
In general, context can be defined as \textit{\quote{any information used to characterize the situation of an entity. An entity is a person, place, or object that is considered relevant to the interaction between a user and an application, including the user and the application themselves, and, by extension, the environment in which the user and applications are embedded. A system is context-aware if it uses context to provide relevant information and/or services to the user, where relevancy depends on the user's task.}} \cite{dey_conceptual_2001, abowd_towards_1999}. In the development of context awareness in augmented reality (AR) or pervasive augmented reality, Grubert et al. \cite{grubert_towards_2017} propose a taxonomy considering context sources (i.e., context factors to which AR systems can adapt), context targets (i.e., content that should be adapted), and context controllers (i.e., how the adaptation should be made). Although they outline high-level categories of context sources (e.g., human, environmental, system factors) and context targets (e.g., system input, system output, system configuration), they do not offer specific guidance for potential capabilities or implementations for context-aware AR systems. While Dey et al. \cite{dey_conceptual_2001} provide guidelines and an architecture for developing context-aware applications through the Context Toolkit, they do not address the various stakeholders' requirements for such systems. Therefore, we first attempt to identify the capabilities for an intelligent wearable context-aware AR system and then adapt the factors and guidelines provided by Dey et al. \cite{dey_conceptual_2001} and Grubert et al. \cite{grubert_towards_2017} to develop such a system.

\newcommand{\PSI}[0]{\textbackslash{psi}} 
The Platform for Situated Intelligence (\PSI{}), developed by Bohus, Andrist, et al. \cite{bohus_platform_2021, andrist_developing_2022}, shares certain high-level objectives and architectural similarities with \TOM{}; both systems utilize timestamped multimodal streaming data from various sensors to facilitate rapid development and research within interactive systems. However, \TOM{} specifically focuses on wearable, user-centered AR applications, necessitating distinct requirements and system capabilities such as user sensing, compared to traditional situated interactive systems targeted by \PSI{}. Moreover, the interactions and interfaces of \TOM{} are optimized to provide intelligent assistance during daily activities while minimizing interference with daily tasks, with the intention of enabling dynamic context- and user-dependent adaptive interfaces.

\subsection{Assistive AR Systems}
\label{sec:related_work:assistive}

Augmented and Mixed Reality (AR/MR) have been used to develop assistive systems that augment users' perception with digitally superimposed content in the physical world \cite{azuma_survey_1997, speicher_what_2019}. These systems have been applied in various domains such as education, training, repair and maintenance, healthcare and medicine, and education due to their advantages in minimizing errors and reducing cognitive load \cite{billinghurst_survey_2015, wang_comprehensive_2016,barsom_systematic_2016}. However, most of these systems are tailored to specific tasks and lack adaptability for various daily activities. 

Several toolkits have been proposed for developing assistive AR systems, yet they are often designed for specialized purposes. For instance, RagRug \cite{fleck_ragrug_2023} and DXR (Data visualizations in eXtended Reality) \cite{sicat_dxr_2019} are designed for situated and immersive analytics, enabling the authoring of context-aware visualizations. 
Project Aria \cite{engel_project_2023} facilitates the collection of egocentric multi-modal data for training ML/AI models in context-aware assistive AR systems but does not offer implementation guidance. 
In contrast, ARGUS (Augmented Reality Guidance and User-modeling System) \cite{castelo_argus_2023}, a visual analytics tool, specializes in multimodal data collection and modeling environmental and performer behaviors. It supports retrospective data analysis through AR sensors and ML models to enhance task guidance. ARGUS aligns partially with our envisioned system, facilitating the generation of large amounts of annotated data for training and developing ML/AI models. These models are designed to enhance decision-making in assistive AR systems or to improve existing systems by visualizing and understanding behaviors. However, as an analytics tool, ARGUS lacks in-built development support for comprehensive end-to-end assistive AR systems (e.g., sensor data integration, context understanding, decision-making for assistance, dynamic feedback delivery via actuators) across various application contexts. This is a gap that our system, \TOM{}, attempts to address.
\added{The very recently introduced SIGMA (Situated Interactive Guidance, Monitoring, and Assistance) \cite{bohus_sigma_2024}, extends \PSI{}, enabling a mature system for (linear) procedural task guidance using MR. While it has advanced 3D scene understanding and visualization capabilities (due to \PSI{}), it lacks system-level capabilities to understand users and primarily supports specific hardware (i.e., HoloLens2), and does not support general-purpose daily tasks by default, which \TOM{} attempts to address.}

With the advancement of AI technologies, new commercial wearable assistive AI systems like Humane Ai Pin \cite{noauthor_ai_2024} and Rabbit R1 \cite{rabbit2023} have emerged, supporting specific activities (e.g., booking taxis, querying objects). 
However, these systems display a limited understanding of users, often restricted to speech interactions, and they do not provide adequate support for developers and researchers in analyzing and visualizing data to enhance assistance or support new activities. In contrast, our system, \TOM{}, aims to understand users' multimodal interactions in context and provide multimodal assistance using AI technologies. Additionally, it attempts to empower developers and researchers to create assistance for various activities, offering robust support for data recording capabilities.


\section{\TOM{}: \TheOtherMe{}}
\label{sec:tom}

\subsection{Envisioned Usage Scenarios}
Consider Jane, who regularly uses \Jerry{}, \textbf{a digital assistant developed using \TOM{}}, in her daily life. \Jerry{} sees what Jane sees, hears what Jane hears, knows her preferences, and understands her emotional and physical conditions.

\paragraph{Scenario} 
Unable to decide on a dish to prepare for herself and her toddler and wishing to try something new, she opens the refrigerator and asks, \quote{Hey, \Jerry{}. Can you suggest a new dish for us?} \Jerry{} scans and identifies the ingredients in the refrigerator, finds possible dishes, and renders three new dish suggestions with their images. Jane finds the second suggestion appealing and inquires about the preparation process. \Jerry{} then guides her through preparing the new dish, providing real-time, step-by-step feedback superimposed in real-world objects.

Later, Jane receives a delivery of a play table set for her toddler, ordered through \Jerry{} during an online browsing session. She notices her toddler's eagerness to assist in assembling the set. Examining the package, she asks, \quote{Hey Tom, can you help me build this?} By identifying the play table set and retrieving instructions, \Jerry{} displays step-by-step virtual instructions superimposed on the physical parts, which Jane follows while involving her toddler. 
Suddenly, her toddler accidentally drops a piece of the set, striking his leg and causing him to cry. Jane becomes panicked. Sensing the situation, \Jerry{} instructs her to remain calm and inspect her toddler's leg. As Jane consoles her child, \Jerry{} assesses the situation and provides first-aid instructions. During the first aid, \Jerry{} asks whether to contact her husband, family doctor, or hospital for further treatment. Upon request, \Jerry{} connects with the family doctor via video call to further observe the toddler's leg.

\subsection{System Capabilities}
\label{sec:system_requirements}
In our quest for an envisioned intelligent wearable assistant, \Jerry{}, we observed that while certain capabilities are supported by existing context-aware and assistive AR/MR systems, a complete integration of these capabilities into a single system is lacking. The Heads-Up Computing Paradigm \cite{zhao_heads_up_2023}, while theoretically supporting our envisioned use cases, does not provide guidance on implementing such a system or the capabilities required to further research optimal Human-AI interactions during daily activities. Drawing from literature, our experience working with AR/MR and AI researchers, and testing assistive Human-AI interfaces (including early prototypes of \TOM{}) with participants and their feedback, we have formulated the following system capabilities. These are categorized based on three major stakeholders' requirements, which, though distinct, have overlapping capabilities.

\paragraph{Just-in-time Assistance for Users}
Users should be able to \textbf{interact} (\RIa{}) with the system (i.e., provide input and receive feedback) naturally and optimally to obtain the desired assistance \cite{hornbaek_what_2017, zhao_heads_up_2023}. Such assistance should be delivered just in time to match the user's current needs or proactively when users have limited knowledge of system capabilities \cite{rhodes_just_time_2000, yang_re_examining_2020, amershi_guidelines_2019}, with minimal interference in the user's ongoing activities while accommodating the user's cognitive capabilities \cite{anderson_survey_2018, mccrickard_attuning_2003}. To achieve this, the system should \textbf{understand the user (\RIb{}) and context (\RIc{})} to provide the most appropriate feedback to support the user's ongoing activities \cite{dey_conceptual_2001, zhao_heads_up_2023}. Such understanding aids in modeling the human and the world to minimize awareness mismatch between user expectations and system feedback and maintaining profiles \cite{schmidt_context_aware_2014}. 

\paragraph{Data Recording and Analysis for Researchers}
To understand user interactions with such a system and to design optimal interactions, researchers need to \textbf{record (\RIIa{}), visualize (\RIIb{}), and analyze (\RIIc{})} the data and develop models \cite{castelo_argus_2023, murray_smith_what_2022,engel_project_2023}. This involves collecting data to support real-time and retrospective observations, training models to predict optimal feedback and analyzing their performance, and understanding the underlying reasons for user and system behaviors \cite{engel_project_2023, oulasvirta_computational_2022, castelo_argus_2023}.

\paragraph{Ease of Development for Developers}
Considering the variety of activities users may engage in and their unique assistance requirements, the system should enable developers to create different assistive features easily. This requires that developers can \textbf{integrate new devices (\RIIIa{})} easily (e.g., sensors to understand new contexts or actuators to provide optimal feedback), \textbf{deploy new assistance and models (\RIIIb{})} (e.g., to predict optimal feedback), and \textbf{access and control current data (\RIIIc{})} (e.g., from existing devices or models).

\subsection{Conceptual Architecture}

\begin{figure*}[hptb]
\centering
\begin{subfigure}{0.35\textwidth}
  \centering
  \includegraphics[width=0.8\linewidth]{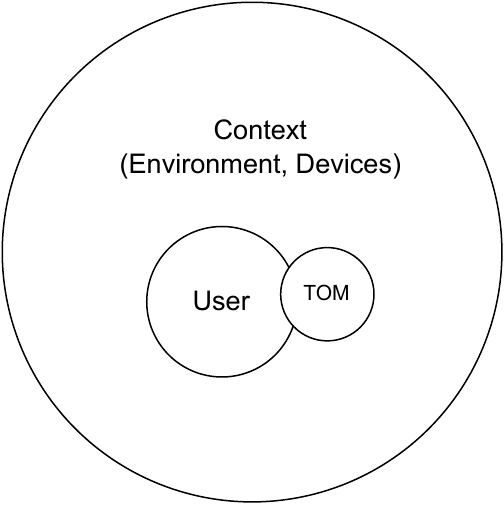}
  \caption{Conceptual entities related to \TOM{}.}
  \label{fig:conceptual_architecture:entities}
\end{subfigure}%
\begin{subfigure}{0.65\textwidth}
  \centering
  \includegraphics[width=0.85\linewidth]{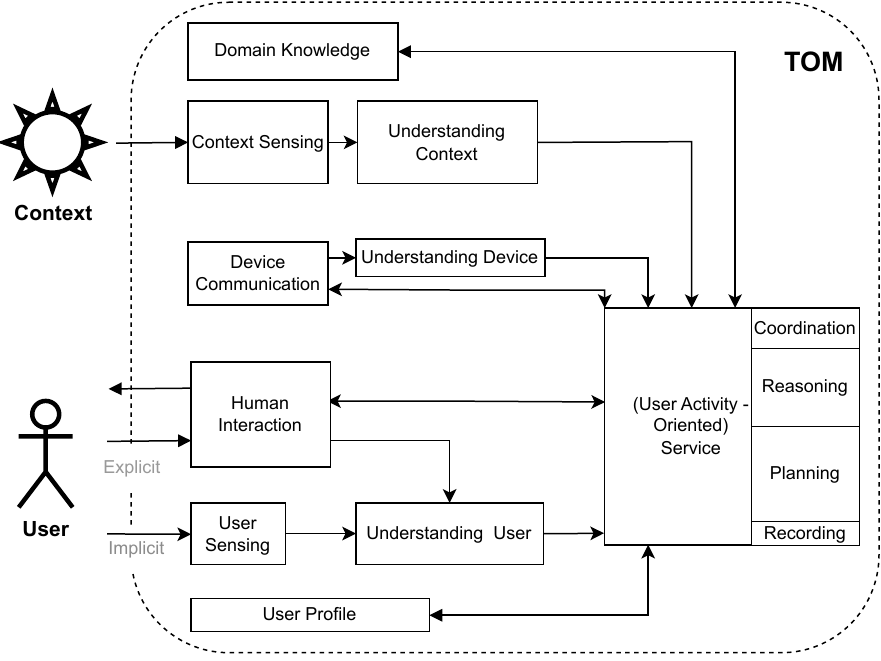}
  \caption{High-level conceptual architecture of \TOM{}.}
  \label{fig:conceptual_architecture:modules}
\end{subfigure}
\caption{Conceptual entities and high-level modules associated with \TOM{}. \added{Arrow directions represent the communication (e.g., data/interaction) flow.}}
\Description{The figure consists of two subfigures: (a) and (b). Subfigure (a) presents the conceptual entities related to TOM, including the Context and the User. Subfigure (b) illustrates the high-level conceptual modules of TOM, encompassing sensing and understanding the user, device, context, human interactions, activity-oriented services for reasoning, planning, and coordination to provide optimal feedback, as well as data recording for analysis.}
\label{fig:conceptual_architecture}
\end{figure*}

To support the above requirements, we consider three main entities: \textit{user} (i.e., the individual receiving assistance), \textit{context} (i.e., the user's perceptual space and associated tasks), and the system, \TOM{}, as shown in Figure~\ref{fig:conceptual_architecture:entities}, following the high-level context sources \cite{grubert_towards_2017}.

Separating the \textit{user} from the \textit{context} enables us to develop user interaction models \cite{zhao_heads_up_2023}. These models sense and understand the user (\RIb{}, e.g., cognitive states, affective states, physical states \cite{grubert_towards_2017}) to provide personalized feedback. Thus, \TOM{} maintains user profiles to cater to individual preferences and capabilities.

Given that daily activities, such as cooking, typically involve both digital (e.g., viewing a recipe) and physical tasks (e.g., selecting the proper portion), \TOM{} offers system-level support to connect the digital world with the physical world by understanding the context (\RIc{}, e.g., physical environment) and utilizing pervasive augmented reality \cite{grubert_towards_2017}. This involves a multi-modal and multifaceted understanding of the environment (e.g., visual and auditory scene understanding, understanding the ongoing activities and associated objects, understanding the relationship between the user and environment, understanding the social context) as well as understanding the devices that facilitate interactions (e.g., device resource availability). 

In terms of input, \TOM{} supports the user's explicit multi-modal inputs (\RIa{}, such as voice and gesture) as well as implicit inputs (\RIa{}, like gaze and physiological data), in addition to processing multi-modal context information.

After understanding the context (e.g., ongoing activity) and user (e.g., intention), \TOM{} activates a context-aware service, employing domain knowledge to generate real-time proactive suggestions through reasoning and planning. 
These suggestions are conveyed to users as multi-sensory feedback (\RIa{}), tailored to their cognitive capacity, including visual, auditory, and/or haptic modalities. The feedback is dynamically updated based on the user's actions; for instance, if the user does not follow a given suggestion, \TOM{} formulates the next appropriate suggestion, considering the user's current status and context, facilitating a closed-loop control system (See Appendix~\ref{appendix:detailed_architecture}-Figure~\ref{fig:detailed_architecture}).

When multiple \TOM{} users are involved in a collaborative activity (e.g., group discussion on an artifact), each \TOM{} system enables multi-agent coordination to complete the collaborative activity optimally.

\subsection{System Architecture}
We develop the following system/physical architecture based on the requirements and envisioned use cases.

\subsubsection{Devices and Technologies}
Following a user-centric approach \cite{zhao_heads_up_2023}, \TOM{} uses wearable devices that align with human input-output channels (e.g., eyes, hands), such as Optical See-Through Head-Mounted Displays (OHMD, Augmented Reality Smart Glasses, e.g., HoloLens2, Nreal Light) and ring-mouses, to support multi-modal interactions (\RIa{}). It also includes everyday wearable devices like smartwatches (e.g., Samsung Galaxy/WearOS, Fitbit) to understand users, smartphones (e.g., Android) to provide familiar interactions, and web browsers for visualizations. 

To understand users, devices, and the environment, and to facilitate reasoning, planning, and coordination, \TOM{} employs AI \cite{rzepka_user_2018} technologies. These include scene understanding, speech recognition, object recognition/tracking, natural language processing, and large language/multimodal models (LLM, LMM). For user feedback, \TOM{} utilizes AR/MR technologies (e.g., OHMD). \TOM{} uses databases (e.g., PostgreSQL, Milvus) for data recording (\RIIa{}), training (\RIIIb{}), and visualization (\RIIb{}). Communication between devices and with external APIs (e.g., ChatGPT) is handled using data communication protocols (e.g., WebSocket, WebRTC, REST API) and wireless mediums (e.g., WiFi, BLE).

\begin{figure*}[hptb]
\centering
\begin{subfigure}{0.5\textwidth}
  \centering
  \includegraphics[width=1\linewidth]{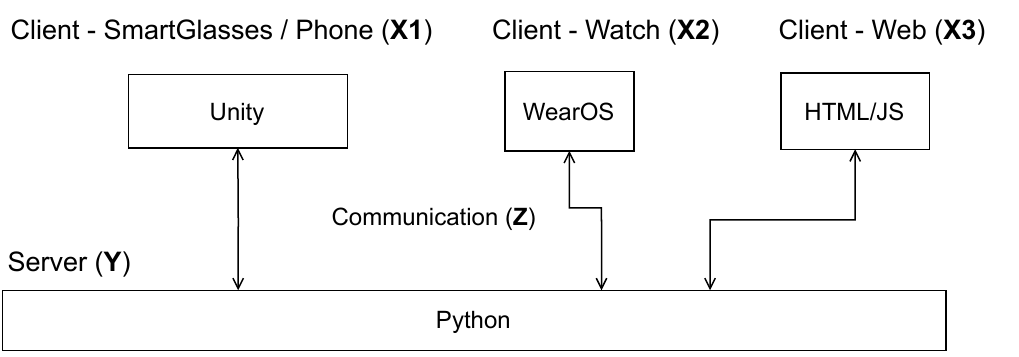}
  \caption{System (physical) architecture.}
  \label{fig:system_architecture:overall}
\end{subfigure}%
\begin{subfigure}{0.22\textwidth}
  \centering
  \includegraphics[width=0.9\linewidth]{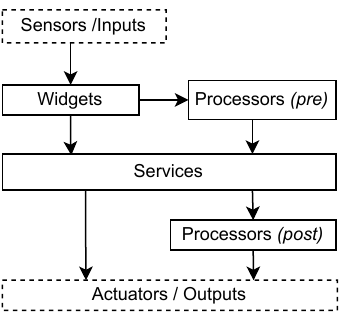}
  \caption{Server architecture.}
  \label{fig:system_architecture:server}
\end{subfigure}%
\begin{subfigure}{0.23\textwidth}
  \centering
  \includegraphics[width=0.9\linewidth]{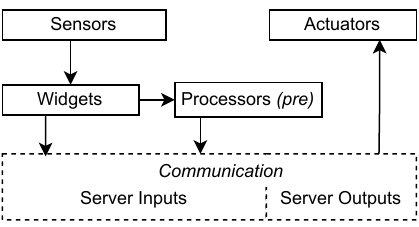}
  \caption{Client architecture.}
  \label{fig:system_architecture:client}
\end{subfigure}%

\caption{System architecture of \TOM{}. Arrow directions represent the data flow. (a) Client-Server architecture with multiple clients (b) Server architecture with multiple layers (c) Client architecture with multiple layers. Dashed boxes indicate the communication between the Server and Clients.}
\Description{The figure comprises three subfigures: (a), (b), and (c). Subfigure (a) displays the overall Client-Server architecture, showing multiple clients such as Smart Glasses, Phones, Watches, and Web clients, communicating with the Server. It also illustrates the technology stack, including Python, Unity, WearOS, and HTML/JS. Subfigure (b) details the Server Architecture, highlighting Widgets that receive Sensor data or Inputs, (Pre-) Processors, Services that process data, and (Post-) Processors that send data to Actuators or Outputs. Subfigure (c) shows the Client Architecture, demonstrating how Sensors connect to the Server via Widgets or Processors, and how Server outputs are transmitted to the Actuators of the Client.}
\label{fig:system_architecture}
\end{figure*}

\subsection{Client-Server Architecture}
Given the limited computational resources of wearable devices \cite{castelo_argus_2023}, \TOM{} is implemented as a client-server architecture, as illustrated in Figure~\ref{fig:system_architecture}. The server hosts services for processing and orchestrating data, supports ML/AI inferences, and provides real-time feedback to clients. Clients, such as OHMDs and smartwatches, send sensor data to the server and display feedback to the user. This separation also allows \TOM{} to be device-agnostic, supporting various OHMDs through the same server (\RIIIa{}).

\paragraph{Server Architecture}
Designed for flexibility and simplicity, the server acts as a one-stop platform for deploying various services optimized for different activities and switching between them as needed. 
Adapting the architecture of the Context Toolkit \cite{dey_conceptual_2001}, the \TOM{} server (Figure~\ref{fig:system_architecture:server}) is implemented with independent components under three distinct layers: Widgets (i.e., components that listen for sensors and receive input data), Processors (i.e., stateless components that process and transform input data from Widgets or output data from Services), and Services (\RIIIb{}, i.e., stateful components that process data from Widgets and/or Processors to produce desired outcomes)\footnote{\TOM{} does not incorporate Interpreters and Aggregators, unlike the Context Toolkit; their functions are managed by either Processors or Services in \TOM{}, minimizing stateful components to enhance testability.}. These layers are interconnected with Sensors/Inputs and Actuators/Outputs linked to Clients. This setup supports the separation of concerns, distributed communication, context storage, and resource discovery. A specialized service, the Context Service, determines the most suitable Service based on current input data (e.g., through explicit user interactions or automatically determined by context data), switching services to support ongoing user activities.

\paragraph{Client Architecture}
As shown in Figure~\ref{fig:system_architecture:client}, Clients mirror the Server's architecture. However, instead of Services, they interact with the Server to stream sensor data and receive real-time feedback for actuators. \added{Time-critical processing can also be implemented in clients (i.e., on-device) to overcome potential latency issues between the client and server.}

\begin{figure*}[hptb]
\centering
\begin{subfigure}{0.44\textwidth}
  \centering
  \includegraphics[width=0.98\linewidth]{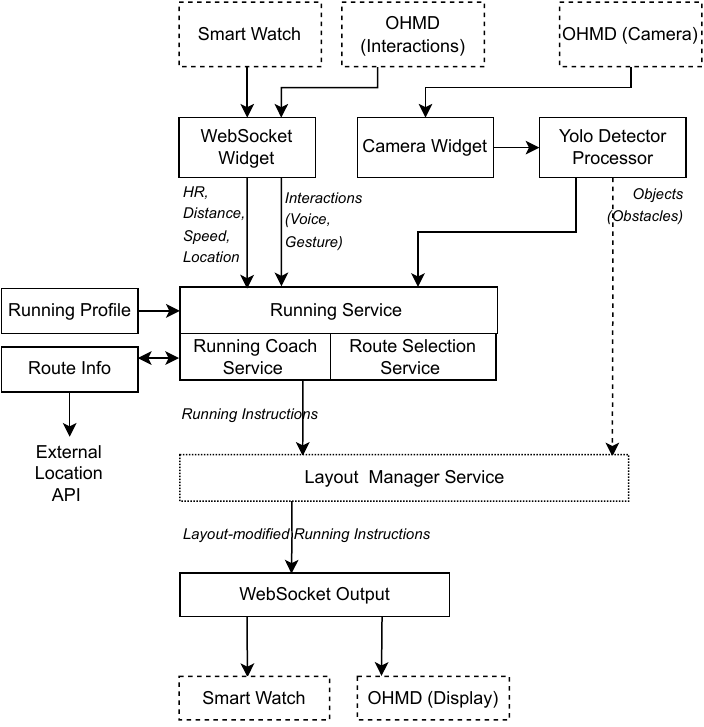}
  \caption{}
  \label{fig:running_service:components}
\end{subfigure}%
\begin{subfigure}{0.56\textwidth}
  \centering
  \includegraphics[width=0.98\linewidth]{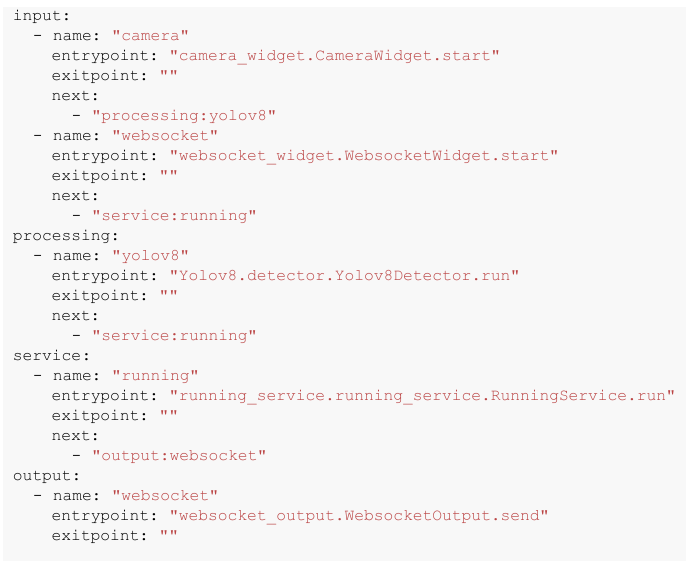}
  \caption{}
  \label{fig:running_service:configuration}
\end{subfigure}

\caption{Running assistance implemented in \TOM{}. (a) System components that enable the running assistance, \RIIIb{}. Dashed-line boxes indicate implemented Client components, solid lines represent implemented Server components, and dotted lines denote Server components under development. (b) The configuration file that controls the data flow, \RIIIc{}. \added{Data is received in one or more components in the Input Layer (e.g., `camera' component) and is sent to the next component as specified in the \textit{next} key (e.g., `yolov8' component in the Processing Layer). This process occurs similarly for all components regardless of the layer, with the \textit{entry point} dictating the method in each component that receives the data from the previous component. The \textit{exit point} then dictates the method for each component, which is called when they should be stopped (e.g., when the context switch indicates the component is no longer required).}}
\Description{The figure consists of two subfigures: (a) and (b). Subfigure (a) depicts the high-level components of the running assistance, including widgets like WebSocket and Camera; processors such as YoloDetector; profiles like Running Profile; domain knowledge like Route Info; services including Running Service and Layout Management Service; and output components like WebSocket. Subfigure (b) presents the configuration file for the Running Service, detailing the data flow, which includes inputs, processors, services, and outputs components, each with a name, entry point, exit point, and next location for data passage.}
\label{fig:running_service}
\end{figure*}

\paragraph{Data Flow and Communication}
Data or messages, tagged with source and time, are transferred between different layers (e.g., Input -> Widget -> Processor -> Service -> Output, Figure~\ref{fig:system_architecture:server}, Appendix~\ref{appendix:detailed_architecture}-Figure\ref{fig:detailed_architecture}) via message channels controlled by configuration files (\RIIIc{}, e.g.,  Figure\ref{fig:running_service:configuration}). This arrangement allows for the reuse of various components across multiple Services and supports distributed communication, thus reducing development efforts. 
Moreover, each component can store the data it handles in a local database for post-analysis (\RIIc{}, e.g., visualization, aggregation) or ML model training (\RIIIb{}). 
The WebSocket protocol is employed for real-time bidirectional communication between Clients and the Server. REST APIs are used for communication with external APIs, while WebRTC and the Real-Time Streaming Protocol (RTSP) facilitate the streaming of real-time video data.

\subsection{Implementation}
The Server (Figure~\ref{fig:system_architecture:overall}-Y) is implemented in Python (3.9), chosen for its extensive user base and support of numerous ML/AI libraries with multiprocessing capabilities. The built-in library, SQLAlchemy, also supports Data Storage.

The OHMD or mobile phone clients (Figure~\ref{fig:system_architecture:overall}-X1) are developed using Unity3D (2021.3) and MRTK\footnote{Mixed Reality Toolkit: \url{https://github.com/Microsoft/MixedRealityToolkit-Unity}} (2.8) to provide AR/MR content. This setup accommodates various devices (e.g., HoloLens2 with UWP, Nreal with Android) and enables mixed reality capabilities with OpenXR support. Other clients, such as the Smartwatch (Figure~\ref{fig:system_architecture:overall}-X2), designed to sense the user, are implemented using WearOS due to its widespread use. Additionally, a web client (Figure~\ref{fig:system_architecture:overall}-X3) is used for visualization, utilizing HTML/VueJS and NodeJS.

For more details, please refer to  \added{\url{https://github.com/TOM-Platform}} and Appendix~\ref{appendix:detailed_architecture}, which details the supported capabilities, devices, technologies, and data.

\section{Demonstration During Daily Activities}
\label{sec:demonstrations}
We have implemented several proof-of-concept services to support daily activities, realizing our vision of an intelligent, wearable, proactive assistant.

\subsection{During Exercise: Running Assistance}
\label{sec:assistance:running}

\paragraph{Scenario}
Jack uses \Jerry{} (implemented using \TOM{}) to assist with his running exercises. He wears an OHMD and a smartwatch (connected to a Server operating on a laptop\footnote{In the future, we plan to run the Server in the cloud}). He initiates his running (speed or distance) training using voice interactions. \Jerry{} provides route options, and he selects one using either voice commands or mid-air gestures. During his run, \Jerry{} provides personalized running coach instructions (e.g., speeding up or slowing down based on his current speed, training plan, and user profile) and proactive suggestions (e.g., encouraging feedback based on duration, alerting about potential dangers like traffic lights based on environment sensing, giving direction cues based on location, indicating waterpoints based on location) using either visual or auditory modality when required (i.e., by default, \Jerry{} will provide only essential details, such as the time, to reduce the display clutter and information overload). At the end of his run, he receives a summary of the exercise.

\begin{figure}[htbp]
\centering
\includegraphics[width=1\linewidth]{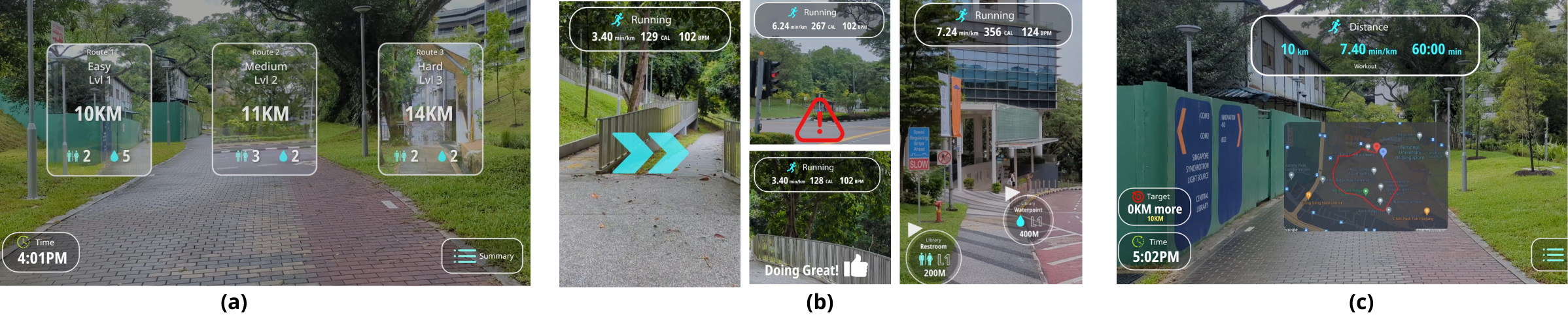}
\caption{The running assistance UI supports voice and mid-air gesture input interactions. (a) The user starts the running assistance and is prompted to select a route. (b) \Jerry{} provides personalized training guidance, proactive feedback on potential dangers or encouragement, and details about water points while running. (c) In the end, \Jerry{} presents the user with a run summary.}
\Description{Three-panel illustration of the running assistance UI. Panel A shows the route selection interface, Panel B displays the in-progress running interface with guidance and feedback, and Panel C showcases the post-run summary screen.}
\label{fig:UI:running}
\end{figure}

\paragraph{System} 
As shown in Figure~\ref{fig:running_service}, the current running assistance is implemented as a Running Service. This service processes sensor data from the smartwatch, user interactions, and the egocentric camera view from the OHMD, and route information from external API to determine the next running coach instruction and provide feedback. Then, the service sends the feedback to the OHMD using a pre-configured display layout (Figure~\ref{fig:UI:running}) designed based on user testing. If the user does not specify certain details required for running (e.g., expected speed), the system uses the user's profile to determine them.

\paragraph{Limitations}
During preliminary user testing, we identified several device and technical limitations. These include impaired visibility of the OHMD's visual feedback in outdoor environments, misrecognition of voice commands due to background noise and user fatigue during running, the OHMD's weight affecting the exercise experience, and instability of visual feedback from frequent head movements (content jumps) \cite{itoh_towards_2021, janaka_glassmessaging_2023}. Additionally, participants requested adaptive user interfaces tailored to their preferences and environment \cite{orlosky_dynamic_2013} for enhanced visibility.

\subsection{During Dining and Shopping: Translation and Querying Assistance}
\label{sec:assistance:dining}

\begin{figure}[htbp]
\centering
\includegraphics[width=1\linewidth]{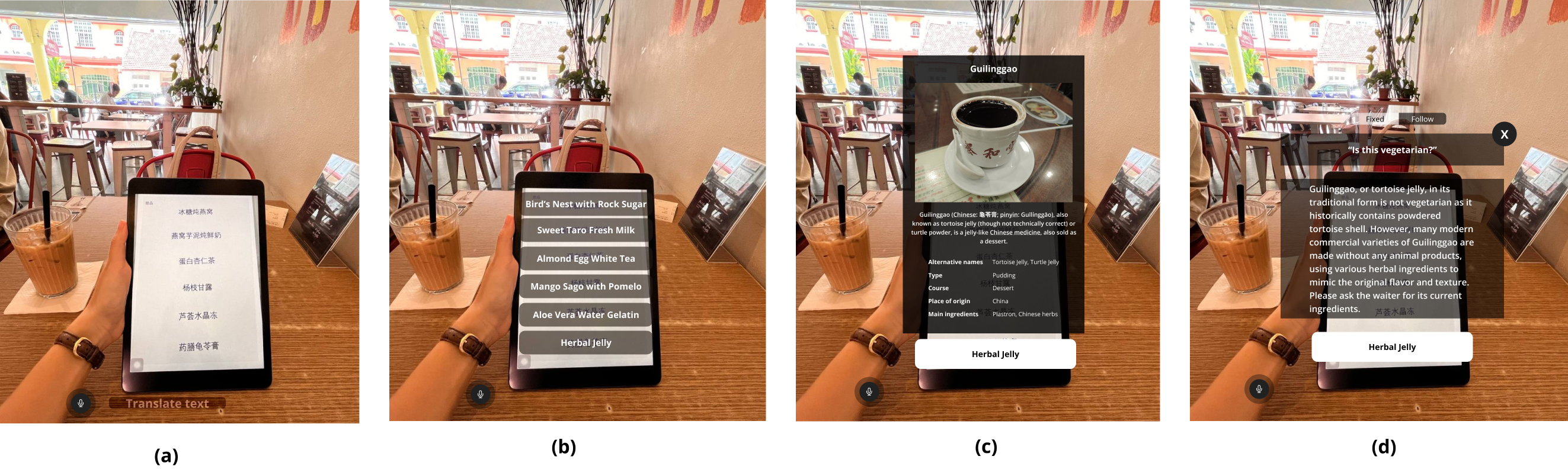}
\caption{The translation assistance UI supports voice, mid-air gesture, and gaze input interactions. (a) The user is prompted for the action they wish to take and chooses the `Translate Text' option. (b) \Jerry{} translates the Mandarin text on the screen into English and overlays the translated information onto the location of the Mandarin text. (c) The user shows interest in `Herbal Jelly' and seeks more general information. (d) The user verbally inquires, \quote{Ok \Jerry{}, is this vegetarian?}}
\Description{Four-panel illustration of the translation assistance UI. Panel A shows the action selection interface. Panel B displays the screen with Mandarin text being translated to English and overlaid. Panel C focuses on 'Herbal Jelly' with additional information being displayed. Panel D depicts the user asking Jerry about the vegetarian status of the item.}
\label{fig:UI:translation}
\end{figure}

\paragraph{Scenario}
Jack visits a new restaurant and discovers that the menu is only in Mandarin, which he cannot understand (See Figure~\ref{fig:UI:translation} for details). He verbally requests \Jerry{}'s assistance, and it displays the translated menu in English, superimposed on the original menu. When he shows prolonged gaze duration on a particular menu item, \Jerry{} automatically displays an image with a brief description. Upon inquiring whether it is vegetarian, \Jerry{} informs him that it contains some non-vegetarian ingredients. Consequently, Jack orders the recommended dish.

After dining, Jack goes to a supermarket. He notices some medical supplements that are different from his usual purchases. Jack asks \Jerry{} about them while looking at (or pointing to) them, and \Jerry{} provides detailed information about them and their usage, assisting him in making an informed decision before purchasing.

\paragraph{System} 
The current translation assistance system uses an OHMD camera to scan and recognize texts, translates them, and employs a Large Language Model (LLM) to adapt them to the local context. Subsequently, \TOM{} displays the translations aligned with the original menu. Based on the user's explicit (e.g., voice) or implicit (e.g., gaze) inputs, \TOM{} presents the menu images and detailed descriptions and verbally answers questions. 
Similarly, the querying assistance system scans and recognizes objects and texts using the OHMD camera, based on the user's explicit interactions (e.g., gaze plus voice, gesture plus voice). It then interacts with an LLM to provide responses to the user's verbal queries.

\paragraph{Limitations}
Similar to the running assistance, additional limitations were noted in the translation and querying assistance. These include the inability to recognize small text in dim lighting, occasional retrieval of incorrect images for specific dishes, and delays (2-8 seconds) in displaying feedback, which is attributable to the response times of external APIs (e.g., ChatGPT, Bing Image Search).



\section{Limitations and Future Improvements}


In addition to the identified technical limitations from specific demonstrations, interaction design challenges surfaced in daily activities. Situational impairments, especially in dynamic environments, restrict certain user interactions (e.g., diminished voice command accuracy in outdoor wind), underscoring the need for methods to support seamless input modality transitions \cite{janaka_glassmessaging_2023}. Inaccuracies in AI-generated suggestions also contribute to user mistrust, necessitating more transparent AI explanations \cite{carroll_why_2022, xu_xair_2023, eiband_bringing_2018}.

Moreover, \TOM{}'s current implementation exhibits limitations. A notable area is the automatic switching of services based on user inputs to optimize service execution for ongoing activities, when multiple services match expected assistance\added{, requiring further research in this area}. Despite \TOM{}'s support for the Large Language Model (LLM) in facilitating human-like conversations and tasks \cite{bubeck_sparks_2023, yang_dawn_2023, brown_language_2020, radford_language_2019}, integrating Large Action Models (LAM) \cite{rabbit2023} could enhance interaction efficiency with external applications and improve user action understanding. Effective live monitoring is available, but \TOM{} needs better visualizations for comprehensive retrospective analysis of long-term user behaviors. Aggregated visualization techniques, similar to ARGUS \cite{castelo_argus_2023}, \added{and retrospective analysis such as PilotAR \cite{janaka_pilotar_2024}} could aid in this. Additionally, advancing the system's grasp of users' cognitive states and their activity correlations requires sophisticated modeling and simulation \cite{murray_smith_what_2022, oulasvirta_computational_2022}. Similarly, facilitating effective multi-agent collaboration while preserving user autonomy in multi-user \TOM{} scenarios remains a significant research challenge.

Finally, developing such systems implicates privacy, security, safety, and ethical challenges. Despite Institutional Review Board (IRB) approval for user studies, real-world deployment raises critical privacy, safety, and social acceptability concerns, considering both users and bystanders \cite{slater_ethics_2020, alallah_performer_2018, denning_situ_2014}. Issues include monitoring and recording users' physical and cognitive states, capturing bystanders' behaviors without consent, securely handling sensitive data, and anonymizing data for aggregate analysis. Although on-device/edge computing provides partial solutions \cite{zhang_data_2018}, the limitations of current devices necessitate further advancements.




\section{Conclusion}

We have presented the anticipated capabilities of developing an intelligent wearable assistive system and introduced \TOM{}, an architecture and open-source implementation  (\added{\url{https://github.com/TOM-Platform}}) that enables researchers and developers to create and analyze assistive applications for supporting daily activities. 
\added{We welcome contributions from the community to expand its supported devices and usage scenarios.}
We envision that \TOM{} will serve as a software platform for researchers and developers to develop innovative, intelligent assistance in various tasks, facilitating human-computer, human-AI, and human-robot interactions. 
Our future plans include extending \TOM{}'s capabilities to enable remote robot interactions, where humans can share information (e.g., intentions) with a remote robot to execute tasks.

\begin{acks}
We would like to express our gratitude to the volunteers who participated in our studies and to the interns of \TOM{} project who developed certain system components, including Teo Yun Yew Jarrett.

This research is supported by the National Research Foundation, Singapore, under its AI Singapore Programme (AISG Award No: AISG2-RP-2020-016). 
The CityU Start-up Grant 9610677 also provides partial support.
Any opinions, findings, conclusions, or recommendations expressed in this material are those of the author(s) and do not reflect the views of the National Research Foundation, Singapore.

\end{acks}

\bibliographystyle{ACM-Reference-Format}
\bibliography{paper/references}


\begin{thebibliography}{44}


\ifx \showCODEN    \undefined \def \showCODEN     #1{\unskip}     \fi
\ifx \showDOI      \undefined \def \showDOI       #1{#1}\fi
\ifx \showISBNx    \undefined \def \showISBNx     #1{\unskip}     \fi
\ifx \showISBNxiii \undefined \def \showISBNxiii  #1{\unskip}     \fi
\ifx \showISSN     \undefined \def \showISSN      #1{\unskip}     \fi
\ifx \showLCCN     \undefined \def \showLCCN      #1{\unskip}     \fi
\ifx \shownote     \undefined \def \shownote      #1{#1}          \fi
\ifx \showarticletitle \undefined \def \showarticletitle #1{#1}   \fi
\ifx \showURL      \undefined \def \showURL       {\relax}        \fi
\providecommand\bibfield[2]{#2}
\providecommand\bibinfo[2]{#2}
\providecommand\natexlab[1]{#1}
\providecommand\showeprint[2][]{arXiv:#2}

\bibitem[noa(2024)]%
        {noauthor_ai_2024}
 \bibinfo{year}{2024}\natexlab{}.
\newblock \bibinfo{title}{Ai {Pin} {Overview}}.
\newblock
\newblock
\urldef\tempurl%
\url{https://hu.ma.ne/aipin}
\showURL{%
\tempurl}


\bibitem[Abowd et~al\mbox{.}(1999)]%
        {abowd_towards_1999}
\bibfield{author}{\bibinfo{person}{Gregory~D. Abowd}, \bibinfo{person}{Anind~K. Dey}, \bibinfo{person}{Peter~J. Brown}, \bibinfo{person}{Nigel Davies}, \bibinfo{person}{Mark Smith}, {and} \bibinfo{person}{Pete Steggles}.} \bibinfo{year}{1999}\natexlab{}.
\newblock \showarticletitle{Towards a {Better} {Understanding} of {Context} and {Context}-{Awareness}}. In \bibinfo{booktitle}{\emph{Handheld and {Ubiquitous} {Computing}}} \emph{(\bibinfo{series}{Lecture {Notes} in {Computer} {Science}})}, \bibfield{editor}{\bibinfo{person}{Hans-W. Gellersen}} (Ed.). \bibinfo{publisher}{Springer}, \bibinfo{address}{Berlin, Heidelberg}, \bibinfo{pages}{304--307}.
\newblock
\showISBNx{978-3-540-48157-7}
\urldef\tempurl%
\url{https://doi.org/10.1007/3-540-48157-5_29}
\showDOI{\tempurl}


\bibitem[Alallah et~al\mbox{.}(2018)]%
        {alallah_performer_2018}
\bibfield{author}{\bibinfo{person}{Fouad Alallah}, \bibinfo{person}{Ali Neshati}, \bibinfo{person}{Yumiko Sakamoto}, \bibinfo{person}{Khalad Hasan}, \bibinfo{person}{Edward Lank}, \bibinfo{person}{Andrea Bunt}, {and} \bibinfo{person}{Pourang Irani}.} \bibinfo{year}{2018}\natexlab{}.
\newblock \showarticletitle{Performer vs. observer: whose comfort level should we consider when examining the social acceptability of input modalities for head-worn display?}. In \bibinfo{booktitle}{\emph{Proceedings of the 24th {ACM} {Symposium} on {Virtual} {Reality} {Software} and {Technology}}} \emph{(\bibinfo{series}{{VRST} '18})}. \bibinfo{publisher}{Association for Computing Machinery}, \bibinfo{address}{New York, NY, USA}, \bibinfo{pages}{1--9}.
\newblock
\showISBNx{978-1-4503-6086-9}
\urldef\tempurl%
\url{https://doi.org/10.1145/3281505.3281541}
\showDOI{\tempurl}


\bibitem[Amershi et~al\mbox{.}(2019)]%
        {amershi_guidelines_2019}
\bibfield{author}{\bibinfo{person}{Saleema Amershi}, \bibinfo{person}{Dan Weld}, \bibinfo{person}{Mihaela Vorvoreanu}, \bibinfo{person}{Adam Fourney}, \bibinfo{person}{Besmira Nushi}, \bibinfo{person}{Penny Collisson}, \bibinfo{person}{Jina Suh}, \bibinfo{person}{Shamsi Iqbal}, \bibinfo{person}{Paul~N. Bennett}, \bibinfo{person}{Kori Inkpen}, \bibinfo{person}{Jaime Teevan}, \bibinfo{person}{Ruth Kikin-Gil}, {and} \bibinfo{person}{Eric Horvitz}.} \bibinfo{year}{2019}\natexlab{}.
\newblock \showarticletitle{Guidelines for {Human}-{AI} {Interaction}}. In \bibinfo{booktitle}{\emph{Proceedings of the 2019 {CHI} {Conference} on {Human} {Factors} in {Computing} {Systems}}} \emph{(\bibinfo{series}{{CHI} '19})}. \bibinfo{publisher}{Association for Computing Machinery}, \bibinfo{address}{New York, NY, USA}, \bibinfo{pages}{1--13}.
\newblock
\showISBNx{978-1-4503-5970-2}
\urldef\tempurl%
\url{https://doi.org/10.1145/3290605.3300233}
\showDOI{\tempurl}


\bibitem[Anderson et~al\mbox{.}(2018)]%
        {anderson_survey_2018}
\bibfield{author}{\bibinfo{person}{Christoph Anderson}, \bibinfo{person}{Isabel Hübener}, \bibinfo{person}{Ann-Kathrin Seipp}, \bibinfo{person}{Sandra Ohly}, \bibinfo{person}{Klaus David}, {and} \bibinfo{person}{Veljko Pejovic}.} \bibinfo{year}{2018}\natexlab{}.
\newblock \showarticletitle{A {Survey} of {Attention} {Management} {Systems} in {Ubiquitous} {Computing} {Environments}}.
\newblock \bibinfo{journal}{\emph{Proceedings of the ACM on Interactive, Mobile, Wearable and Ubiquitous Technologies}} \bibinfo{volume}{2}, \bibinfo{number}{2} (\bibinfo{date}{July} \bibinfo{year}{2018}), \bibinfo{pages}{1--27}.
\newblock
\showISSN{2474-9567, 2474-9567}
\urldef\tempurl%
\url{https://doi.org/10.1145/3214261}
\showDOI{\tempurl}


\bibitem[Andrist et~al\mbox{.}(2022)]%
        {andrist_developing_2022}
\bibfield{author}{\bibinfo{person}{Sean Andrist}, \bibinfo{person}{Dan Bohus}, \bibinfo{person}{Ashley Feniello}, {and} \bibinfo{person}{Nick Saw}.} \bibinfo{year}{2022}\natexlab{}.
\newblock \showarticletitle{Developing {Mixed} {Reality} {Applications} with {Platform} for {Situated} {Intelligence}}. In \bibinfo{booktitle}{\emph{2022 {IEEE} {Conference} on {Virtual} {Reality} and {3D} {User} {Interfaces} {Abstracts} and {Workshops} ({VRW})}}. \bibinfo{pages}{48--50}.
\newblock
\urldef\tempurl%
\url{https://doi.org/10.1109/VRW55335.2022.00018}
\showDOI{\tempurl}


\bibitem[Azuma(1997)]%
        {azuma_survey_1997}
\bibfield{author}{\bibinfo{person}{Ronald~T Azuma}.} \bibinfo{year}{1997}\natexlab{}.
\newblock \showarticletitle{A {Survey} of {Augmented} {Reality}}.
\newblock \bibinfo{journal}{\emph{Presence: Teleoperators and Virtual Environments}} \bibinfo{volume}{6}, \bibinfo{number}{4} (\bibinfo{date}{Aug.} \bibinfo{year}{1997}), \bibinfo{pages}{355--385}.
\newblock
\showISSN{1054-7460}
\urldef\tempurl%
\url{https://doi.org/10.1162/pres.1997.6.4.355}
\showDOI{\tempurl}


\bibitem[Barsom et~al\mbox{.}(2016)]%
        {barsom_systematic_2016}
\bibfield{author}{\bibinfo{person}{E.~Z. Barsom}, \bibinfo{person}{M. Graafland}, {and} \bibinfo{person}{M.~P. Schijven}.} \bibinfo{year}{2016}\natexlab{}.
\newblock \showarticletitle{Systematic review on the effectiveness of augmented reality applications in medical training}.
\newblock \bibinfo{journal}{\emph{Surgical Endoscopy}} \bibinfo{volume}{30}, \bibinfo{number}{10} (\bibinfo{date}{Oct.} \bibinfo{year}{2016}), \bibinfo{pages}{4174--4183}.
\newblock
\showISSN{1432-2218}
\urldef\tempurl%
\url{https://doi.org/10.1007/s00464-016-4800-6}
\showDOI{\tempurl}


\bibitem[Billinghurst et~al\mbox{.}(2015)]%
        {billinghurst_survey_2015}
\bibfield{author}{\bibinfo{person}{Mark Billinghurst}, \bibinfo{person}{Adrian Clark}, {and} \bibinfo{person}{Gun Lee}.} \bibinfo{year}{2015}\natexlab{}.
\newblock \showarticletitle{A {Survey} of {Augmented} {Reality}}.
\newblock \bibinfo{journal}{\emph{Foundations and Trends® in Human–Computer Interaction}} \bibinfo{volume}{8}, \bibinfo{number}{2-3} (\bibinfo{date}{March} \bibinfo{year}{2015}), \bibinfo{pages}{73--272}.
\newblock
\showISSN{1551-3955, 1551-3963}
\urldef\tempurl%
\url{https://doi.org/10.1561/1100000049}
\showDOI{\tempurl}


\bibitem[Bohus et~al\mbox{.}(2021)]%
        {bohus_platform_2021}
\bibfield{author}{\bibinfo{person}{Dan Bohus}, \bibinfo{person}{Sean Andrist}, \bibinfo{person}{Ashley Feniello}, \bibinfo{person}{Nick Saw}, \bibinfo{person}{Mihai Jalobeanu}, \bibinfo{person}{Patrick Sweeney}, \bibinfo{person}{Anne~Loomis Thompson}, {and} \bibinfo{person}{Eric Horvitz}.} \bibinfo{year}{2021}\natexlab{}.
\newblock \bibinfo{title}{Platform for {Situated} {Intelligence}}.
\newblock
\newblock
\urldef\tempurl%
\url{https://doi.org/10.48550/arXiv.2103.15975}
\showDOI{\tempurl}
\newblock
\shownote{arXiv:2103.15975 [cs]}.


\bibitem[Bohus et~al\mbox{.}(2024)]%
        {bohus_sigma_2024}
\bibfield{author}{\bibinfo{person}{Dan Bohus}, \bibinfo{person}{Sean Andrist}, \bibinfo{person}{Nick Saw}, \bibinfo{person}{Ann Paradiso}, \bibinfo{person}{Ishani Chakraborty}, {and} \bibinfo{person}{Mahdi Rad}.} \bibinfo{year}{2024}\natexlab{}.
\newblock \bibinfo{title}{{SIGMA}: {An} {Open}-{Source} {Interactive} {System} for {Mixed}-{Reality} {Task} {Assistance} {Research}}.
\newblock
\newblock
\urldef\tempurl%
\url{https://arxiv.org/abs/2405.13035v1}
\showURL{%
\tempurl}


\bibitem[Brown et~al\mbox{.}(2020)]%
        {brown_language_2020}
\bibfield{author}{\bibinfo{person}{Tom~B. Brown}, \bibinfo{person}{Benjamin Mann}, \bibinfo{person}{Nick Ryder}, \bibinfo{person}{Melanie Subbiah}, \bibinfo{person}{Jared Kaplan}, \bibinfo{person}{Prafulla Dhariwal}, \bibinfo{person}{Arvind Neelakantan}, \bibinfo{person}{Pranav Shyam}, \bibinfo{person}{Girish Sastry}, \bibinfo{person}{Amanda Askell}, \bibinfo{person}{Sandhini Agarwal}, \bibinfo{person}{Ariel Herbert-Voss}, \bibinfo{person}{Gretchen Krueger}, \bibinfo{person}{Tom Henighan}, \bibinfo{person}{Rewon Child}, \bibinfo{person}{Aditya Ramesh}, \bibinfo{person}{Daniel~M. Ziegler}, \bibinfo{person}{Jeffrey Wu}, \bibinfo{person}{Clemens Winter}, \bibinfo{person}{Christopher Hesse}, \bibinfo{person}{Mark Chen}, \bibinfo{person}{Eric Sigler}, \bibinfo{person}{Mateusz Litwin}, \bibinfo{person}{Scott Gray}, \bibinfo{person}{Benjamin Chess}, \bibinfo{person}{Jack Clark}, \bibinfo{person}{Christopher Berner}, \bibinfo{person}{Sam McCandlish}, \bibinfo{person}{Alec Radford}, \bibinfo{person}{Ilya Sutskever},
  {and} \bibinfo{person}{Dario Amodei}.} \bibinfo{year}{2020}\natexlab{}.
\newblock \bibinfo{title}{Language {Models} are {Few}-{Shot} {Learners}}.
\newblock
\newblock
\urldef\tempurl%
\url{https://doi.org/10.48550/arXiv.2005.14165}
\showDOI{\tempurl}


\bibitem[Bubeck et~al\mbox{.}(2023)]%
        {bubeck_sparks_2023}
\bibfield{author}{\bibinfo{person}{Sébastien Bubeck}, \bibinfo{person}{Varun Chandrasekaran}, \bibinfo{person}{Ronen Eldan}, \bibinfo{person}{Johannes Gehrke}, \bibinfo{person}{Eric Horvitz}, \bibinfo{person}{Ece Kamar}, \bibinfo{person}{Peter Lee}, \bibinfo{person}{Yin~Tat Lee}, \bibinfo{person}{Yuanzhi Li}, \bibinfo{person}{Scott Lundberg}, \bibinfo{person}{Harsha Nori}, \bibinfo{person}{Hamid Palangi}, \bibinfo{person}{Marco~Tulio Ribeiro}, {and} \bibinfo{person}{Yi Zhang}.} \bibinfo{year}{2023}\natexlab{}.
\newblock \bibinfo{title}{Sparks of {Artificial} {General} {Intelligence}: {Early} experiments with {GPT}-4}.
\newblock
\newblock
\urldef\tempurl%
\url{https://doi.org/10.48550/arXiv.2303.12712}
\showDOI{\tempurl}


\bibitem[Carroll(2022)]%
        {carroll_why_2022}
\bibfield{author}{\bibinfo{person}{John~M. Carroll}.} \bibinfo{year}{2022}\natexlab{}.
\newblock \showarticletitle{Why should humans trust {AI}?}
\newblock \bibinfo{journal}{\emph{Interactions}} \bibinfo{volume}{29}, \bibinfo{number}{4} (\bibinfo{date}{June} \bibinfo{year}{2022}), \bibinfo{pages}{73--77}.
\newblock
\showISSN{1072-5520}
\urldef\tempurl%
\url{https://doi.org/10.1145/3538392}
\showDOI{\tempurl}


\bibitem[Castelo et~al\mbox{.}(2023)]%
        {castelo_argus_2023}
\bibfield{author}{\bibinfo{person}{Sonia Castelo}, \bibinfo{person}{Joao Rulff}, \bibinfo{person}{Erin McGowan}, \bibinfo{person}{Bea Steers}, \bibinfo{person}{Guande Wu}, \bibinfo{person}{Shaoyu Chen}, \bibinfo{person}{Iran Roman}, \bibinfo{person}{Roque Lopez}, \bibinfo{person}{Ethan Brewer}, \bibinfo{person}{Chen Zhao}, \bibinfo{person}{Jing Qian}, \bibinfo{person}{Kyunghyun Cho}, \bibinfo{person}{He He}, \bibinfo{person}{Qi Sun}, \bibinfo{person}{Huy Vo}, \bibinfo{person}{Juan Bello}, \bibinfo{person}{Michael Krone}, {and} \bibinfo{person}{Claudio Silva}.} \bibinfo{year}{2023}\natexlab{}.
\newblock \bibinfo{title}{{ARGUS}: {Visualization} of {AI}-{Assisted} {Task} {Guidance} in {AR}}.
\newblock
\newblock
\urldef\tempurl%
\url{https://doi.org/10.48550/arXiv.2308.06246}
\showDOI{\tempurl}


\bibitem[Denning et~al\mbox{.}(2014)]%
        {denning_situ_2014}
\bibfield{author}{\bibinfo{person}{Tamara Denning}, \bibinfo{person}{Zakariya Dehlawi}, {and} \bibinfo{person}{Tadayoshi Kohno}.} \bibinfo{year}{2014}\natexlab{}.
\newblock \showarticletitle{In situ with bystanders of augmented reality glasses: perspectives on recording and privacy-mediating technologies}. In \bibinfo{booktitle}{\emph{Proceedings of the 32nd annual {ACM} conference on {Human} factors in computing systems - {CHI} '14}}. \bibinfo{publisher}{ACM Press}, \bibinfo{address}{Toronto, Ontario, Canada}, \bibinfo{pages}{2377--2386}.
\newblock
\showISBNx{978-1-4503-2473-1}
\urldef\tempurl%
\url{https://doi.org/10.1145/2556288.2557352}
\showDOI{\tempurl}


\bibitem[Dey et~al\mbox{.}(2001)]%
        {dey_conceptual_2001}
\bibfield{author}{\bibinfo{person}{Anind~K. Dey}, \bibinfo{person}{Gregory~D. Abowd}, {and} \bibinfo{person}{Daniel Salber}.} \bibinfo{year}{2001}\natexlab{}.
\newblock \showarticletitle{A {Conceptual} {Framework} and a {Toolkit} for {Supporting} the {Rapid} {Prototyping} of {Context}-{Aware} {Applications}}.
\newblock \bibinfo{journal}{\emph{Human–Computer Interaction}} \bibinfo{volume}{16}, \bibinfo{number}{2-4} (\bibinfo{date}{Dec.} \bibinfo{year}{2001}), \bibinfo{pages}{97--166}.
\newblock
\showISSN{0737-0024}
\urldef\tempurl%
\url{https://doi.org/10.1207/S15327051HCI16234_02}
\showDOI{\tempurl}


\bibitem[Eiband et~al\mbox{.}(2018)]%
        {eiband_bringing_2018}
\bibfield{author}{\bibinfo{person}{Malin Eiband}, \bibinfo{person}{Hanna Schneider}, \bibinfo{person}{Mark Bilandzic}, \bibinfo{person}{Julian Fazekas-Con}, \bibinfo{person}{Mareike Haug}, {and} \bibinfo{person}{Heinrich Hussmann}.} \bibinfo{year}{2018}\natexlab{}.
\newblock \showarticletitle{Bringing {Transparency} {Design} into {Practice}}. In \bibinfo{booktitle}{\emph{23rd {International} {Conference} on {Intelligent} {User} {Interfaces}}} \emph{(\bibinfo{series}{{IUI} '18})}. \bibinfo{publisher}{Association for Computing Machinery}, \bibinfo{address}{New York, NY, USA}, \bibinfo{pages}{211--223}.
\newblock
\showISBNx{978-1-4503-4945-1}
\urldef\tempurl%
\url{https://doi.org/10.1145/3172944.3172961}
\showDOI{\tempurl}


\bibitem[Engel et~al\mbox{.}(2023)]%
        {engel_project_2023}
\bibfield{author}{\bibinfo{person}{Jakob Engel}, \bibinfo{person}{Kiran Somasundaram}, \bibinfo{person}{Michael Goesele}, \bibinfo{person}{Albert Sun}, \bibinfo{person}{Alexander Gamino}, \bibinfo{person}{Andrew Turner}, \bibinfo{person}{Arjang Talattof}, \bibinfo{person}{Arnie Yuan}, \bibinfo{person}{Bilal Souti}, \bibinfo{person}{Brighid Meredith}, \bibinfo{person}{Cheng Peng}, \bibinfo{person}{Chris Sweeney}, \bibinfo{person}{Cole Wilson}, \bibinfo{person}{Dan Barnes}, \bibinfo{person}{Daniel DeTone}, \bibinfo{person}{David Caruso}, \bibinfo{person}{Derek Valleroy}, \bibinfo{person}{Dinesh Ginjupalli}, \bibinfo{person}{Duncan Frost}, \bibinfo{person}{Edward Miller}, \bibinfo{person}{Elias Mueggler}, \bibinfo{person}{Evgeniy Oleinik}, \bibinfo{person}{Fan Zhang}, \bibinfo{person}{Guruprasad Somasundaram}, \bibinfo{person}{Gustavo Solaira}, \bibinfo{person}{Harry Lanaras}, \bibinfo{person}{Henry Howard-Jenkins}, \bibinfo{person}{Huixuan Tang}, \bibinfo{person}{Hyo~Jin Kim}, \bibinfo{person}{Jaime Rivera},
  \bibinfo{person}{Ji Luo}, \bibinfo{person}{Jing Dong}, \bibinfo{person}{Julian Straub}, \bibinfo{person}{Kevin Bailey}, \bibinfo{person}{Kevin Eckenhoff}, \bibinfo{person}{Lingni Ma}, \bibinfo{person}{Luis Pesqueira}, \bibinfo{person}{Mark Schwesinger}, \bibinfo{person}{Maurizio Monge}, \bibinfo{person}{Nan Yang}, \bibinfo{person}{Nick Charron}, \bibinfo{person}{Nikhil Raina}, \bibinfo{person}{Omkar Parkhi}, \bibinfo{person}{Peter Borschowa}, \bibinfo{person}{Pierre Moulon}, \bibinfo{person}{Prince Gupta}, \bibinfo{person}{Raul Mur-Artal}, \bibinfo{person}{Robbie Pennington}, \bibinfo{person}{Sachin Kulkarni}, \bibinfo{person}{Sagar Miglani}, \bibinfo{person}{Santosh Gondi}, \bibinfo{person}{Saransh Solanki}, \bibinfo{person}{Sean Diener}, \bibinfo{person}{Shangyi Cheng}, \bibinfo{person}{Simon Green}, \bibinfo{person}{Steve Saarinen}, \bibinfo{person}{Suvam Patra}, \bibinfo{person}{Tassos Mourikis}, \bibinfo{person}{Thomas Whelan}, \bibinfo{person}{Tripti Singh}, \bibinfo{person}{Vasileios Balntas},
  \bibinfo{person}{Vijay Baiyya}, \bibinfo{person}{Wilson Dreewes}, \bibinfo{person}{Xiaqing Pan}, \bibinfo{person}{Yang Lou}, \bibinfo{person}{Yipu Zhao}, \bibinfo{person}{Yusuf Mansour}, \bibinfo{person}{Yuyang Zou}, \bibinfo{person}{Zhaoyang Lv}, \bibinfo{person}{Zijian Wang}, \bibinfo{person}{Mingfei Yan}, \bibinfo{person}{Carl Ren}, \bibinfo{person}{Renzo De~Nardi}, {and} \bibinfo{person}{Richard Newcombe}.} \bibinfo{year}{2023}\natexlab{}.
\newblock \bibinfo{title}{Project {Aria}: {A} {New} {Tool} for {Egocentric} {Multi}-{Modal} {AI} {Research}}.
\newblock
\newblock
\urldef\tempurl%
\url{https://doi.org/10.48550/arXiv.2308.13561}
\showDOI{\tempurl}


\bibitem[Fleck et~al\mbox{.}(2023)]%
        {fleck_ragrug_2023}
\bibfield{author}{\bibinfo{person}{Philipp Fleck}, \bibinfo{person}{Aimée~Sousa Calepso}, \bibinfo{person}{Sebastian Hubenschmid}, \bibinfo{person}{Michael Sedlmair}, {and} \bibinfo{person}{Dieter Schmalstieg}.} \bibinfo{year}{2023}\natexlab{}.
\newblock \showarticletitle{{RagRug}: {A} {Toolkit} for {Situated} {Analytics}}.
\newblock \bibinfo{journal}{\emph{IEEE Transactions on Visualization and Computer Graphics}} \bibinfo{volume}{29}, \bibinfo{number}{7} (\bibinfo{date}{July} \bibinfo{year}{2023}), \bibinfo{pages}{3281--3297}.
\newblock
\showISSN{1941-0506}
\urldef\tempurl%
\url{https://doi.org/10.1109/TVCG.2022.3157058}
\showDOI{\tempurl}
\newblock
\shownote{Conference Name: IEEE Transactions on Visualization and Computer Graphics}.


\bibitem[Grubert et~al\mbox{.}(2017)]%
        {grubert_towards_2017}
\bibfield{author}{\bibinfo{person}{Jens Grubert}, \bibinfo{person}{Tobias Langlotz}, \bibinfo{person}{Stefanie Zollmann}, {and} \bibinfo{person}{Holger Regenbrecht}.} \bibinfo{year}{2017}\natexlab{}.
\newblock \showarticletitle{Towards {Pervasive} {Augmented} {Reality}: {Context}-{Awareness} in {Augmented} {Reality}}.
\newblock \bibinfo{journal}{\emph{IEEE Transactions on Visualization and Computer Graphics}} \bibinfo{volume}{23}, \bibinfo{number}{6} (\bibinfo{date}{June} \bibinfo{year}{2017}), \bibinfo{pages}{1706--1724}.
\newblock
\showISSN{1941-0506}
\urldef\tempurl%
\url{https://doi.org/10.1109/TVCG.2016.2543720}
\showDOI{\tempurl}


\bibitem[Hornbæk and Oulasvirta(2017)]%
        {hornbaek_what_2017}
\bibfield{author}{\bibinfo{person}{Kasper Hornbæk} {and} \bibinfo{person}{Antti Oulasvirta}.} \bibinfo{year}{2017}\natexlab{}.
\newblock \showarticletitle{What {Is} {Interaction}?}. In \bibinfo{booktitle}{\emph{Proceedings of the 2017 {CHI} {Conference} on {Human} {Factors} in {Computing} {Systems}}}. \bibinfo{publisher}{ACM}, \bibinfo{address}{Denver Colorado USA}, \bibinfo{pages}{5040--5052}.
\newblock
\showISBNx{978-1-4503-4655-9}
\urldef\tempurl%
\url{https://doi.org/10.1145/3025453.3025765}
\showDOI{\tempurl}


\bibitem[Itoh et~al\mbox{.}(2021)]%
        {itoh_towards_2021}
\bibfield{author}{\bibinfo{person}{Yuta Itoh}, \bibinfo{person}{Tobias Langlotz}, \bibinfo{person}{Jonathan Sutton}, {and} \bibinfo{person}{Alexander Plopski}.} \bibinfo{year}{2021}\natexlab{}.
\newblock \showarticletitle{Towards {Indistinguishable} {Augmented} {Reality}: {A} {Survey} on {Optical} {See}-through {Head}-mounted {Displays}}.
\newblock \bibinfo{journal}{\emph{Comput. Surveys}} \bibinfo{volume}{54}, \bibinfo{number}{6} (\bibinfo{date}{July} \bibinfo{year}{2021}), \bibinfo{pages}{120:1--120:36}.
\newblock
\showISSN{0360-0300}
\urldef\tempurl%
\url{https://doi.org/10.1145/3453157}
\showDOI{\tempurl}


\bibitem[Janaka et~al\mbox{.}(2024)]%
        {janaka_pilotar_2024}
\bibfield{author}{\bibinfo{person}{Nuwan Janaka}, \bibinfo{person}{Runze Cai}, \bibinfo{person}{Ashwin Ram}, \bibinfo{person}{Lin Zhu}, \bibinfo{person}{Shengdong Zhao}, {and} \bibinfo{person}{Yong Kai~Qi}.} \bibinfo{year}{2024}\natexlab{}.
\newblock \showarticletitle{{PilotAR}: {Streamlining} {Pilot} {Studies} with {OHMDs} from {Concept} to {Insight}}.
\newblock \bibinfo{journal}{\emph{Proceedings of the ACM on Interactive, Mobile, Wearable and Ubiquitous Technologies}} (\bibinfo{date}{Sept.} \bibinfo{year}{2024}).
\newblock
\urldef\tempurl%
\url{https://doi.org/10.1145/3678576}
\showDOI{\tempurl}


\bibitem[Janaka et~al\mbox{.}(2023)]%
        {janaka_glassmessaging_2023}
\bibfield{author}{\bibinfo{person}{Nuwan Janaka}, \bibinfo{person}{Jie Gao}, \bibinfo{person}{Lin Zhu}, \bibinfo{person}{Shengdong Zhao}, \bibinfo{person}{Lan Lyu}, \bibinfo{person}{Peisen Xu}, \bibinfo{person}{Maximilian Nabokow}, \bibinfo{person}{Silang Wang}, {and} \bibinfo{person}{Yanch Ong}.} \bibinfo{year}{2023}\natexlab{}.
\newblock \showarticletitle{{GlassMessaging}: {Towards} {Ubiquitous} {Messaging} {Using} {OHMDs}}.
\newblock \bibinfo{journal}{\emph{Proceedings of the ACM on Interactive, Mobile, Wearable and Ubiquitous Technologies}} \bibinfo{volume}{7}, \bibinfo{number}{3} (\bibinfo{date}{Sept.} \bibinfo{year}{2023}), \bibinfo{pages}{100:1--100:32}.
\newblock
\urldef\tempurl%
\url{https://doi.org/10.1145/3610931}
\showDOI{\tempurl}


\bibitem[McCrickard and Chewar(2003)]%
        {mccrickard_attuning_2003}
\bibfield{author}{\bibinfo{person}{D.~Scott McCrickard} {and} \bibinfo{person}{C.~M. Chewar}.} \bibinfo{year}{2003}\natexlab{}.
\newblock \showarticletitle{Attuning notification design to user goals and attention costs}.
\newblock \bibinfo{journal}{\emph{Commun. ACM}} \bibinfo{volume}{46}, \bibinfo{number}{3} (\bibinfo{date}{March} \bibinfo{year}{2003}), \bibinfo{pages}{67}.
\newblock
\showISSN{00010782}
\urldef\tempurl%
\url{https://doi.org/10.1145/636772.636800}
\showDOI{\tempurl}


\bibitem[Murray-Smith et~al\mbox{.}(2022)]%
        {murray_smith_what_2022}
\bibfield{author}{\bibinfo{person}{Roderick Murray-Smith}, \bibinfo{person}{Antti Oulasvirta}, \bibinfo{person}{Andrew Howes}, \bibinfo{person}{Jörg Müller}, \bibinfo{person}{Aleksi Ikkala}, \bibinfo{person}{Miroslav Bachinski}, \bibinfo{person}{Arthur Fleig}, \bibinfo{person}{Florian Fischer}, {and} \bibinfo{person}{Markus Klar}.} \bibinfo{year}{2022}\natexlab{}.
\newblock \showarticletitle{What simulation can do for {HCI} research}.
\newblock \bibinfo{journal}{\emph{Interactions}} \bibinfo{volume}{29}, \bibinfo{number}{6} (\bibinfo{date}{Nov.} \bibinfo{year}{2022}), \bibinfo{pages}{48--53}.
\newblock
\showISSN{1072-5520}
\urldef\tempurl%
\url{https://doi.org/10.1145/3564038}
\showDOI{\tempurl}


\bibitem[Orlosky et~al\mbox{.}(2013)]%
        {orlosky_dynamic_2013}
\bibfield{author}{\bibinfo{person}{Jason Orlosky}, \bibinfo{person}{Kiyoshi Kiyokawa}, {and} \bibinfo{person}{Haruo Takemura}.} \bibinfo{year}{2013}\natexlab{}.
\newblock \showarticletitle{Dynamic text management for see-through wearable and heads-up display systems}. In \bibinfo{booktitle}{\emph{Proceedings of the 2013 international conference on {Intelligent} user interfaces - {IUI} '13}}. \bibinfo{publisher}{ACM Press}, \bibinfo{address}{Santa Monica, California, USA}, \bibinfo{pages}{363}.
\newblock
\showISBNx{978-1-4503-1965-2}
\urldef\tempurl%
\url{https://doi.org/10.1145/2449396.2449443}
\showDOI{\tempurl}


\bibitem[Oulasvirta et~al\mbox{.}(2022)]%
        {oulasvirta_computational_2022}
\bibfield{author}{\bibinfo{person}{Antti Oulasvirta}, \bibinfo{person}{Jussi P.~P. Jokinen}, {and} \bibinfo{person}{Andrew Howes}.} \bibinfo{year}{2022}\natexlab{}.
\newblock \showarticletitle{Computational {Rationality} as a {Theory} of {Interaction}}. In \bibinfo{booktitle}{\emph{Proceedings of the 2022 {CHI} {Conference} on {Human} {Factors} in {Computing} {Systems}}} \emph{(\bibinfo{series}{{CHI} '22})}. \bibinfo{publisher}{Association for Computing Machinery}, \bibinfo{address}{New York, NY, USA}, \bibinfo{pages}{1--14}.
\newblock
\showISBNx{978-1-4503-9157-3}
\urldef\tempurl%
\url{https://doi.org/10.1145/3491102.3517739}
\showDOI{\tempurl}


\bibitem[rabbit~research team(2023)]%
        {rabbit2023}
\bibfield{author}{\bibinfo{person}{rabbit~research team}.} \bibinfo{year}{2023}\natexlab{}.
\newblock \bibinfo{title}{Learning human actions on computer applications}.
\newblock
\newblock
\urldef\tempurl%
\url{https://rabbit.tech/research}
\showURL{%
\tempurl}


\bibitem[Radford et~al\mbox{.}(2019)]%
        {radford_language_2019}
\bibfield{author}{\bibinfo{person}{Alec Radford}, \bibinfo{person}{Jeff Wu}, \bibinfo{person}{Rewon Child}, \bibinfo{person}{D. Luan}, \bibinfo{person}{Dario Amodei}, {and} \bibinfo{person}{Ilya Sutskever}.} \bibinfo{year}{2019}\natexlab{}.
\newblock \showarticletitle{Language {Models} are {Unsupervised} {Multitask} {Learners}}.
\newblock
\urldef\tempurl%
\url{https://api.semanticscholar.org/CorpusID:160025533}
\showURL{%
\tempurl}


\bibitem[Rhodes and Maes(2000)]%
        {rhodes_just_time_2000}
\bibfield{author}{\bibinfo{person}{B.~J. Rhodes} {and} \bibinfo{person}{P. Maes}.} \bibinfo{year}{2000}\natexlab{}.
\newblock \showarticletitle{Just-in-time information retrieval agents}.
\newblock \bibinfo{journal}{\emph{IBM Systems Journal}} \bibinfo{volume}{39}, \bibinfo{number}{3.4} (\bibinfo{year}{2000}), \bibinfo{pages}{685--704}.
\newblock
\showISSN{0018-8670}
\urldef\tempurl%
\url{https://doi.org/10.1147/sj.393.0685}
\showDOI{\tempurl}


\bibitem[Rzepka and Berger(2018)]%
        {rzepka_user_2018}
\bibfield{author}{\bibinfo{person}{Christine Rzepka} {and} \bibinfo{person}{Benedikt Berger}.} \bibinfo{year}{2018}\natexlab{}.
\newblock \showarticletitle{User {Interaction} with {AI}-enabled {Systems}: {A} {Systematic} {Review} of {IS} {Research}}.
\newblock \bibinfo{journal}{\emph{ICIS 2018 Proceedings}} (\bibinfo{date}{Dec.} \bibinfo{year}{2018}).
\newblock
\urldef\tempurl%
\url{https://aisel.aisnet.org/icis2018/general/Presentations/7}
\showURL{%
\tempurl}


\bibitem[Schmidt(2014)]%
        {schmidt_context_aware_2014}
\bibfield{author}{\bibinfo{person}{Albrecht Schmidt}.} \bibinfo{year}{2014}\natexlab{}.
\newblock \bibinfo{booktitle}{\emph{Context-{Aware} {Computing}}}.
\newblock
\urldef\tempurl%
\url{https://www.interaction-design.org/literature/book/the-encyclopedia-of-human-computer-interaction-2nd-ed/context-aware-computing-context-awareness-context-aware-user-interfaces-and-implicit-interaction}
\showURL{%
\tempurl}


\bibitem[Sicat et~al\mbox{.}(2019)]%
        {sicat_dxr_2019}
\bibfield{author}{\bibinfo{person}{Ronell Sicat}, \bibinfo{person}{Jiabao Li}, \bibinfo{person}{Junyoung Choi}, \bibinfo{person}{Maxime Cordeil}, \bibinfo{person}{Won-Ki Jeong}, \bibinfo{person}{Benjamin Bach}, {and} \bibinfo{person}{Hanspeter Pfister}.} \bibinfo{year}{2019}\natexlab{}.
\newblock \showarticletitle{{DXR}: {A} {Toolkit} for {Building} {Immersive} {Data} {Visualizations}}.
\newblock \bibinfo{journal}{\emph{IEEE Transactions on Visualization and Computer Graphics}} \bibinfo{volume}{25}, \bibinfo{number}{1} (\bibinfo{date}{Jan.} \bibinfo{year}{2019}), \bibinfo{pages}{715--725}.
\newblock
\showISSN{1941-0506}
\urldef\tempurl%
\url{https://doi.org/10.1109/TVCG.2018.2865152}
\showDOI{\tempurl}
\newblock
\shownote{Conference Name: IEEE Transactions on Visualization and Computer Graphics}.


\bibitem[Slater et~al\mbox{.}(2020)]%
        {slater_ethics_2020}
\bibfield{author}{\bibinfo{person}{Mel Slater}, \bibinfo{person}{Cristina Gonzalez-Liencres}, \bibinfo{person}{Patrick Haggard}, \bibinfo{person}{Charlotte Vinkers}, \bibinfo{person}{Rebecca Gregory-Clarke}, \bibinfo{person}{Steve Jelley}, \bibinfo{person}{Zillah Watson}, \bibinfo{person}{Graham Breen}, \bibinfo{person}{Raz Schwarz}, \bibinfo{person}{William Steptoe}, \bibinfo{person}{Dalila Szostak}, \bibinfo{person}{Shivashankar Halan}, \bibinfo{person}{Deborah Fox}, {and} \bibinfo{person}{Jeremy Silver}.} \bibinfo{year}{2020}\natexlab{}.
\newblock \showarticletitle{The {Ethics} of {Realism} in {Virtual} and {Augmented} {Reality}}.
\newblock \bibinfo{journal}{\emph{Frontiers in Virtual Reality}}  \bibinfo{volume}{1} (\bibinfo{year}{2020}).
\newblock
\showISSN{2673-4192}
\urldef\tempurl%
\url{https://doi.org/10.3389/frvir.2020.00001}
\showDOI{\tempurl}


\bibitem[Speicher et~al\mbox{.}(2019)]%
        {speicher_what_2019}
\bibfield{author}{\bibinfo{person}{Maximilian Speicher}, \bibinfo{person}{Brian~D. Hall}, {and} \bibinfo{person}{Michael Nebeling}.} \bibinfo{year}{2019}\natexlab{}.
\newblock \showarticletitle{What is {Mixed} {Reality}?}. In \bibinfo{booktitle}{\emph{Proceedings of the 2019 {CHI} {Conference} on {Human} {Factors} in {Computing} {Systems}}} \emph{(\bibinfo{series}{{CHI} '19})}. \bibinfo{publisher}{Association for Computing Machinery}, \bibinfo{address}{New York, NY, USA}, \bibinfo{pages}{1--15}.
\newblock
\showISBNx{978-1-4503-5970-2}
\urldef\tempurl%
\url{https://doi.org/10.1145/3290605.3300767}
\showDOI{\tempurl}


\bibitem[Victor(2018)]%
        {victor_dynamicland_2018}
\bibfield{author}{\bibinfo{person}{Bret Victor}.} \bibinfo{year}{2018}\natexlab{}.
\newblock \bibinfo{title}{Dynamicland}.
\newblock
\newblock
\urldef\tempurl%
\url{https://dynamicland.org/}
\showURL{%
\tempurl}


\bibitem[Wang et~al\mbox{.}(2016)]%
        {wang_comprehensive_2016}
\bibfield{author}{\bibinfo{person}{X. Wang}, \bibinfo{person}{S.~K. Ong}, {and} \bibinfo{person}{A.~Y.~C. Nee}.} \bibinfo{year}{2016}\natexlab{}.
\newblock \showarticletitle{A comprehensive survey of augmented reality assembly research}.
\newblock \bibinfo{journal}{\emph{Advances in Manufacturing}} \bibinfo{volume}{4}, \bibinfo{number}{1} (\bibinfo{date}{March} \bibinfo{year}{2016}), \bibinfo{pages}{1--22}.
\newblock
\showISSN{2195-3597}
\urldef\tempurl%
\url{https://doi.org/10.1007/s40436-015-0131-4}
\showDOI{\tempurl}


\bibitem[Xu et~al\mbox{.}(2023)]%
        {xu_xair_2023}
\bibfield{author}{\bibinfo{person}{Xuhai Xu}, \bibinfo{person}{Anna Yu}, \bibinfo{person}{Tanya~R. Jonker}, \bibinfo{person}{Kashyap Todi}, \bibinfo{person}{Feiyu Lu}, \bibinfo{person}{Xun Qian}, \bibinfo{person}{João~Marcelo Evangelista~Belo}, \bibinfo{person}{Tianyi Wang}, \bibinfo{person}{Michelle Li}, \bibinfo{person}{Aran Mun}, \bibinfo{person}{Te-Yen Wu}, \bibinfo{person}{Junxiao Shen}, \bibinfo{person}{Ting Zhang}, \bibinfo{person}{Narine Kokhlikyan}, \bibinfo{person}{Fulton Wang}, \bibinfo{person}{Paul Sorenson}, \bibinfo{person}{Sophie Kim}, {and} \bibinfo{person}{Hrvoje Benko}.} \bibinfo{year}{2023}\natexlab{}.
\newblock \showarticletitle{{XAIR}: {A} {Framework} of {Explainable} {AI} in {Augmented} {Reality}}. In \bibinfo{booktitle}{\emph{Proceedings of the 2023 {CHI} {Conference} on {Human} {Factors} in {Computing} {Systems}}} \emph{(\bibinfo{series}{{CHI} '23})}. \bibinfo{publisher}{Association for Computing Machinery}, \bibinfo{address}{New York, NY, USA}, \bibinfo{pages}{1--30}.
\newblock
\showISBNx{978-1-4503-9421-5}
\urldef\tempurl%
\url{https://doi.org/10.1145/3544548.3581500}
\showDOI{\tempurl}


\bibitem[Yang et~al\mbox{.}(2020)]%
        {yang_re_examining_2020}
\bibfield{author}{\bibinfo{person}{Qian Yang}, \bibinfo{person}{Aaron Steinfeld}, \bibinfo{person}{Carolyn Rosé}, {and} \bibinfo{person}{John Zimmerman}.} \bibinfo{year}{2020}\natexlab{}.
\newblock \showarticletitle{Re-examining {Whether}, {Why}, and {How} {Human}-{AI} {Interaction} {Is} {Uniquely} {Difficult} to {Design}}. In \bibinfo{booktitle}{\emph{Proceedings of the 2020 {CHI} {Conference} on {Human} {Factors} in {Computing} {Systems}}} \emph{(\bibinfo{series}{{CHI} '20})}. \bibinfo{publisher}{Association for Computing Machinery}, \bibinfo{address}{New York, NY, USA}, \bibinfo{pages}{1--13}.
\newblock
\showISBNx{978-1-4503-6708-0}
\urldef\tempurl%
\url{https://doi.org/10.1145/3313831.3376301}
\showDOI{\tempurl}


\bibitem[Yang et~al\mbox{.}(2023)]%
        {yang_dawn_2023}
\bibfield{author}{\bibinfo{person}{Zhengyuan Yang}, \bibinfo{person}{Linjie Li}, \bibinfo{person}{Kevin Lin}, \bibinfo{person}{Jianfeng Wang}, \bibinfo{person}{Chung-Ching Lin}, \bibinfo{person}{Zicheng Liu}, {and} \bibinfo{person}{Lijuan Wang}.} \bibinfo{year}{2023}\natexlab{}.
\newblock \bibinfo{title}{The {Dawn} of {LMMs}: {Preliminary} {Explorations} with {GPT}-{4V}(ision)}.
\newblock
\newblock
\urldef\tempurl%
\url{https://doi.org/10.48550/arXiv.2309.17421}
\showDOI{\tempurl}


\bibitem[Zhang et~al\mbox{.}(2018)]%
        {zhang_data_2018}
\bibfield{author}{\bibinfo{person}{Jiale Zhang}, \bibinfo{person}{Bing Chen}, \bibinfo{person}{Yanchao Zhao}, \bibinfo{person}{Xiang Cheng}, {and} \bibinfo{person}{Feng Hu}.} \bibinfo{year}{2018}\natexlab{}.
\newblock \showarticletitle{Data {Security} and {Privacy}-{Preserving} in {Edge} {Computing} {Paradigm}: {Survey} and {Open} {Issues}}.
\newblock \bibinfo{journal}{\emph{IEEE Access}}  \bibinfo{volume}{6} (\bibinfo{year}{2018}), \bibinfo{pages}{18209--18237}.
\newblock
\showISSN{2169-3536}
\urldef\tempurl%
\url{https://doi.org/10.1109/ACCESS.2018.2820162}
\showDOI{\tempurl}


\bibitem[Zhao et~al\mbox{.}(2023)]%
        {zhao_heads_up_2023}
\bibfield{author}{\bibinfo{person}{Shengdong Zhao}, \bibinfo{person}{Felicia Tan}, {and} \bibinfo{person}{Katherine Fennedy}.} \bibinfo{year}{2023}\natexlab{}.
\newblock \showarticletitle{Heads-{Up} {Computing} {Moving} {Beyond} the {Device}-{Centered} {Paradigm}}.
\newblock \bibinfo{journal}{\emph{Commun. ACM}} \bibinfo{volume}{66}, \bibinfo{number}{9} (\bibinfo{date}{Aug.} \bibinfo{year}{2023}), \bibinfo{pages}{56--63}.
\newblock
\showISSN{0001-0782}
\urldef\tempurl%
\url{https://doi.org/10.1145/3571722}
\showDOI{\tempurl}


\end{thebibliography}

\appendix

\section{Detailed System Architecture}
\label{appendix:detailed_architecture}

\begin{figure}[hptb]
\centering
\includegraphics[width=1\linewidth]{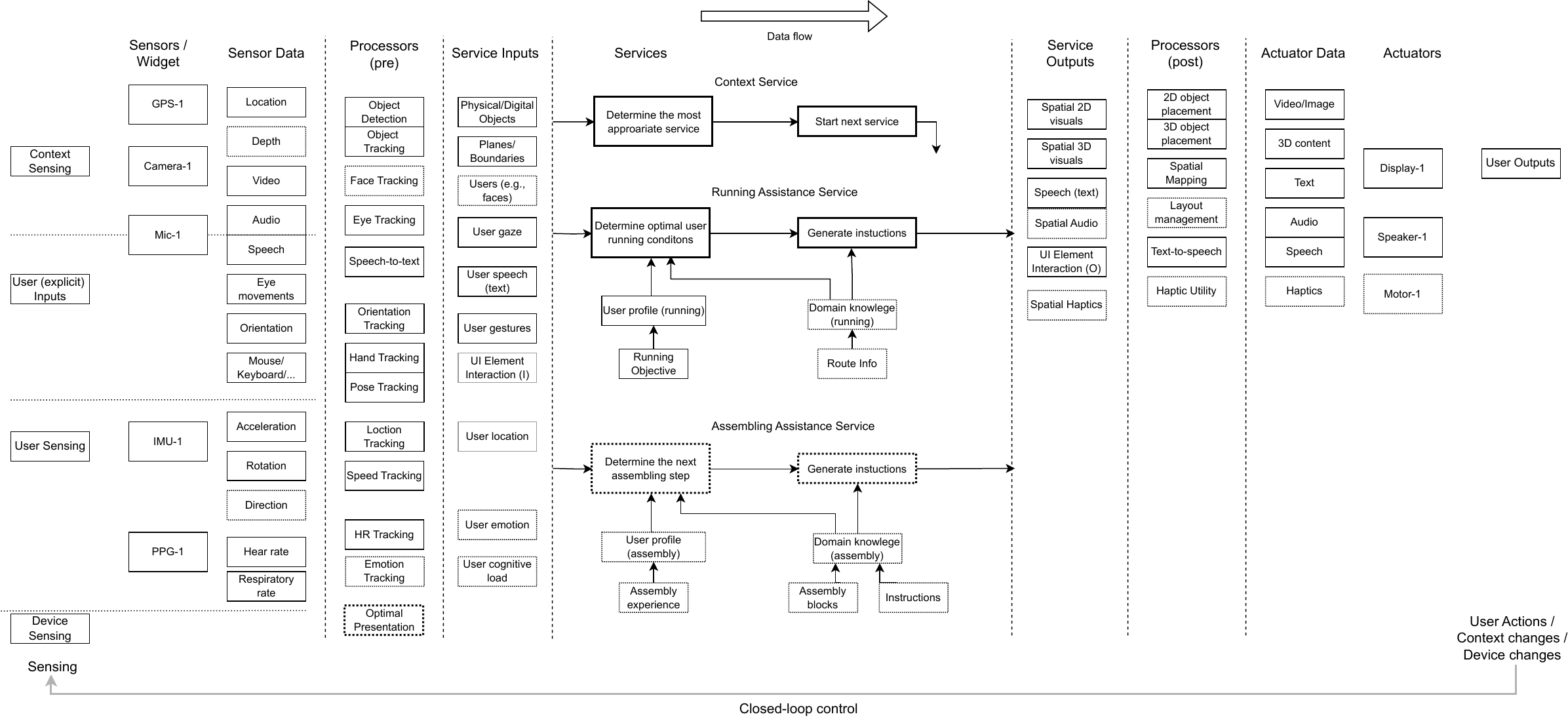}
\caption{A snapshot of the system architecture and the data flow between high-level components. Solid-lined boxes indicate implemented components, and dotted lines represent components in development.}
\Description{The figure illustrates the detailed system architecture and the data flow among high-level components. It starts with Sensors/Widgets capturing sensor data, which is then processed by (Pre-)Processors. This data flows into Service Inputs, leading to various Services that generate Service Outputs. These outputs are further processed by (Post-)Processors, eventually reaching Actuators. The diagram visually represents a closed-loop control system, achieved by continuously sensing the user and context, and adjusting system feedback based on user actions. The flow is depicted through a series of interconnected boxes and arrows, with solid lines for existing components and dotted lines for those under development.}
\label{fig:detailed_architecture}
\end{figure}

Figure~\ref{fig:detailed_architecture} depicts the high-level components of the system architecture.
The implemented Server currently supports three distinct types of Services: running assistance, providing speed/distance training with directional support; learning assistance, enabling inquiries about objects in the frame selected through gestures and/or gaze; and translation assistance, which facilitates text translation within the frame.

Table \ref{tab:system:technologies} provides an overview of the technologies that support the system's current capabilities and 
Table~\ref{tab:system:data} illustrates the various client data supporting the system's capabilities. These data sources include Ring Mouse Controllers connected to OHMDs, Gesture and Gaze Detection through OHMDs, and User Input via the Touch Screen on Android Phones.

\begin{table*}[hptb]
\caption{Technologies used within the current \TOM{}. For the latest supported technologies, refer to the \href{https://github.com/TOM-Platform}{TOM-Platform}}
\label{tab:system:technologies}
\scalebox{0.65}{
\begin{tabular}{@{}lllllll@{}}
\toprule
\multirow{2}{*}{\textbf{Capability}} &
  \multirow{2}{*}{\textbf{Type}} &
  \textbf{Server} &
  \multicolumn{4}{l}{\textbf{Client}} \\ \cmidrule(l){3-7} 
 &
   &
  \textbf{Python Server (Laptop)} &
  \textbf{HoloLens 2} &
  \textbf{XReal Light} &
  \textbf{Web Client} &
  \textbf{Android Phone} \\ \midrule
\multirow{3}{*}{\textbf{Context Understanding}} &
  Visual &
  GoogleCloudVision, Yolov8 &
  - &
  - &
  - &
  MLKit \\
 &
  Audio &
  MediaPipe &
  - &
  - &
  - &
  - \\
 &
  Spatial &
  GoogleCloudVision &
  SLAM (inbuilt) &
  SLAM (inbuilt) &
  - &
  - \\ \midrule
\textbf{User Understanding} &
  Physiological &
  Fitbit API &
  - &
  - &
  - &
  WearOS Watch \\ \midrule
\multirow{3}{*}{\textbf{Interactions}} &
  Visual &
  Bing Image Search API &
  - &
  - &
  Graphs, Charts &
  - \\
 &
  Audio &
  \begin{tabular}[c]{@{}l@{}}Speech-to-Text (Whisper),\\ Text-to-Speech\end{tabular} &
  \begin{tabular}[c]{@{}l@{}}Speech-to-Text (Azura),\\ Text-to-Speech\end{tabular} &
  - &
  - &
  \begin{tabular}[c]{@{}l@{}}Speech-to-Text (Android),\\ Text-to-Speech\end{tabular} \\
 &
  Other &
  - &
  \begin{tabular}[c]{@{}l@{}}Hand Tracking (inbuilt),\\ Gaze Tracking (inbuilt)\end{tabular} &
  Hand Tracking (inbuilt) &
  Visualizations &
  Touch \\ \midrule
\multirow{3}{*}{\textbf{Communication}} &
  APIs &
  REST &
  - &
  - &
  - &
  - \\
 &
  Data Stream &
  Websocket &
  Websocket &
  Websocket, TCPSocket &
  Websocket, TCPSocket &
  TCPSocket \\
 &
  Video Stream &
  WebRTC, RTSP &
  - &
  WebRTC &
  WebRTC &
  WebRTC \\ \midrule
\textbf{Assistance} &
  
  \begin{tabular}[c]{@{}l@{}}Queries,\\Memory \end{tabular} &
  \begin{tabular}[c]{@{}l@{}}ChatGPT, Claude, \\ Local LLM (TinyLlama) \\ Encoding (Clip, ImageBind) \\ Vector Database (Milvus) \end{tabular} &
  - &
  - &
  - &
  - \\ \bottomrule
\end{tabular}
}
\Description{This table outlines the technologies utilized in the system. It comprises five rows, each labeled with a different aspect of the system: Context Understanding, User Understanding, Interactions, Communication, and Assistance. The table also includes column headers for Server (Python) and five Clients (HoloLens2, XReal Light, Web Client, Android Phone), indicating the technologies implemented in each. The arrangement of the table provides a clear depiction of how various technologies are applied across different components of the system.}
\end{table*}

\begin{table*}[hptb]
\caption{Data from currently supported Clients. For the latest supported devices/capabilities, refer to the \url{https://github.com/TOM-Platform}{TOM-Platform}}
\label{tab:system:data}
\scalebox{0.8}{
\begin{tabular}{@{}lllllll@{}}
\toprule
\multirow{2}{*}{\textbf{Data}} &
  \multirow{2}{*}{\textbf{Type}} &
  \multicolumn{5}{c}{\textbf{Client}} \\ \cmidrule(l){3-7} 
 &
   &
  \textbf{HoloLens2} &
  \textbf{XReal Light} &
  \textbf{WearOS Watch} &
  \textbf{Web Client} &
  \textbf{Android Phone} \\ \midrule
\multirow{3}{*}{\textbf{Context Understanding}} &
  Visual &
  Video &
  Video &
  - &
  - &
  Video \\
 &
  Auditory &
  Audio &
  Audio &
  - &
  - &
  Audio \\
 &
  Spatial &
  WorldMesh &
  WorldMesh &
  Location (GPS) &
  - &
  Location (GPS) \\ \midrule
\multirow{2}{*}{\textbf{User Understanding}} &
  Physiological &
  - &
  - &
  Heart Rate &
  - &
  - \\
 &
  Physical &
  - &
  - &
  Speed, Calories &
  - &
  - \\ \midrule
\multirow{2}{*}{\textbf{Interactions (Output)}} &
  Visual &
  \begin{tabular}[c]{@{}l@{}}Text, Image,\\ Video, 3D Object\end{tabular} &
  \begin{tabular}[c]{@{}l@{}}Text, Image,\\ Video, 3D Object\end{tabular} &
  \begin{tabular}[c]{@{}l@{}}Text, Image,\\ 2D Object\end{tabular} &
  \begin{tabular}[c]{@{}l@{}}Text, Images,\\ Video\end{tabular} &
  \begin{tabular}[c]{@{}l@{}}Text, Image,\\ Video, 3D Object\end{tabular} \\
 &
  Auditory &
  Audio, Text &
  Audio, Text &
  Audio &
  Audio &
  Audio, Text \\ \midrule
\multirow{4}{*}{\textbf{Interactions (Input)}} &
  Voice &
  Audio (Speech) &
  Audio (Speech) &
  - &
  - &
  Audio (Speech) \\
 &
  Gesture &
  Hand, Finger &
  Hand, Finger &
  - &
  - &
  - \\
 &
  Gaze &
  3D Gaze, Gaze Collision &
  - &
  - &
  - &
  - \\
 &
  Controller &
  2D Press &
  3D Press &
  - &
  - &
  2D Touch \\ \bottomrule
\end{tabular}
}
\Description{This table details the types of data related to Context Understanding, User Understanding, Interaction Outputs, and Interaction Inputs used in the system. It is structured into categories across rows and specifies the corresponding client technologies across columns, including HoloLens2, XReal Light, WearOS Watch, Web Client, and Android Phone. The table enumerates various data types, such as Visual, Auditory, and Spatial for context understanding, and different modes of interaction like Voice, Gesture, and Gaze.}
\end{table*}

\end{document}